\documentclass[11pt]{article}
\usepackage[utf8]{inputenc}
\usepackage{booktabs}
\usepackage{amsmath}
\usepackage{amssymb}
\usepackage[numbers,sort&compress]{natbib}
\usepackage{comment}
\usepackage{float}
\usepackage{placeins}
\usepackage{soul}

\usepackage{caption}
\usepackage{xcolor}
\usepackage[margin = 2.5cm]{geometry}
\usepackage{graphicx}
\usepackage[hyperfootnotes = false, colorlinks = true, linkcolor = blue, citecolor = purple]{hyperref}
\usepackage{subcaption}
\usepackage{dsfont}
\usepackage{tikz}

\usepackage{fancyhdr}
\parskip=\medskipamount
\newgeometry{margin=2cm}

\definecolor{dgreen}{rgb}{0,0.5,0}
\definecolor{dpink}{rgb}{1,0.3,0.3}
\definecolor{darkblue}{rgb}{0,0,0.6}
\definecolor{purple}{rgb}{0.4,.2,0.7}
\definecolor{avocadogreen}{RGB}{86,130,3}

\newcommand{\smref}[1]{Appendix~\ref{#1}}
\newcommand{\smrefs}[1]{Appendices~#1}
\newcommand{\maineqref}[1]{Eq.~\eqref{#1}}
\newcommand{\mainref}[1]{Fig.~\ref{#1}}

\newcommand{\smsecref}[1]{Sec.~\ref{#1}}
\newcommand{\smfigref}[1]{Fig.~\ref{#1}}

\newcommand{\convergencefigwidth}{0.72\linewidth}
\newenvironment{figurewide}{\begin{figure}[tbp]}{\end{figure}}
\newenvironment{figurewidetop}{\begin{figure}[t!]}{\end{figure}}
\newenvironment{figuremain}{\begin{figure}[h]}{\end{figure}}
\newenvironment{tablewide}{\begin{table}[tbp]}{\end{table}}
\newcommand{\functionalfigwidth}{0.60\linewidth}
\newenvironment{functionalfigure}{\begin{figure}[t!]}{\end{figure}}
\newcommand{\placefunctionalfigureearly}{}

\begin{document}
\renewcommand*{\thefootnote}{\fnsymbol{footnote}}

\thispagestyle{empty}
\begin{center}
    ~\vspace{5mm}
{\LARGE\bfseries
Bootstrapping 3D Conformal Field Theories\\\vspace{0.1cm} with Product Analytic Functionals}
%{\LARGE \bf %Carving Out the Space of 3D %Three-Dimensional 
 % Bootstrapping 3D Conformal Field Theories\\ \vspace{0.25cm} with Analytic Functionals}
 
   \vspace{0.5in}
\begingroup\large
   {\bf Kausik Ghosh\footnote{kau.rock91@gmail.com} $\,$ \it{and} $\,$ \bf Zechuan Zheng\footnote{zechuan.zheng.phy@gmail.com}}
\par\endgroup
    \vspace{0.5in}

\begingroup\sffamily\footnotesize
    \textsuperscript{$\star$}Department of Mathematics, King's College London, Strand, London WC2R 2LS, United Kingdom\\
   \textsuperscript{$\dagger$}Perimeter Institute for Theoretical Physics, Waterloo, ON N2L 2Y5, Canada
 \par\endgroup
    \vspace{0.5in}

\end{center}

\vspace{0.3in}

\vspace{0.3in}

\begin{abstract}
%We introduce a bootstrap method for three-dimensional conformal field theories based on analytic functionals and show that it sharply improves the convergence of numerical bounds. Our construction uses one-dimensional analytic functionals as building blocks for higher-dimensional crossing equations, providing a new alternative to the standard derivative basis. We benchmark the method in single-correlator gap maximization problems, focusing on scalar and spin-2 channels. At comparable truncation size, analytic functionals give substantially stronger bounds, often matching results that require much higher derivative order in the conventional approach. The improvement is most striking at large external dimension, where derivative-based bounds converge slowly and our bounds reveal new sharp features in the space of allowed CFT data. This makes analytic functionals a promising tool for bootstrapping conformal gauge theories, where the presence of heavy external operators poses a major challenge for conventional methods.

We introduce a bootstrap method for three-dimensional conformal
field theories based on product analytic functionals. The construction combines
one-dimensional analytic functionals with dimensional reduction and provides
an efficient alternative to the standard derivative basis. We benchmark the
method in single correlator gap maximization problems for scalar and spin-two
operators. At comparable basis size, the product functional basis gives
stronger bounds and converges more rapidly, with the improvement becoming
especially pronounced at large external dimension. Setting external dimension to that of the three-dimensional Ising spin operator, we substantially sharpen the single correlator upper bound on the leading scalar dimension. We also sharpen the Nakayama–Ohtsuki necessary condition for critical points accessible by tuning a single parameter. In the spin-two
problem, we find a new kink near $\Delta_\phi\simeq4.16$, accompanied by a
reorganization of the extremal spectrum. We also observe plateau structures at
larger external dimension in both gap-maximization problems. Our results make product analytic functionals a promising tool for conformal gauge theories with heavy external operators and for probing the flat-space limit of holographic correlators; accessing both of these scenarios is challenging for conventional methods.

\end{abstract}

\thispagestyle{empty}

\clearpage

\setcounter{tocdepth}{3}

\tableofcontents
\renewcommand*{\thefootnote}{\arabic{footnote}}
\setcounter{footnote}{0}

% ---------------- Main text ----------------
\section{Introduction}

Since its modern revival \cite{Rattazzi:2008pe}, the conformal bootstrap has
become a powerful nonperturbative method for charting the space of conformal
field theories (CFTs). Starting only from conformal symmetry, unitarity, and
crossing symmetry, it constrains consistent CFT data without reference to a
Lagrangian. In three dimensions, this program revealed a prominent kink\footnote{referred to as the ``Ising kink'' throughout the paper.} in
the bound on the leading scalar operator
\cite{El-Showk:2012cjh,El-Showk:2014dwa}, located remarkably close to the three-dimensional
Ising CFT.
Mixed-correlator semidefinite programming
\cite{Kos:2014bka,Simmons-Duffin:2015qma} subsequently produced precision
islands \cite{Kos:2016ysd,Poland:2018epd,Liu:2023elz}, the sharpest determinations of Ising
critical data \cite{Chang:2024whx}, and precise constraints for
theories with global symmetry
\cite{Chester:2019ifh,Erramilli:2022kgp,He:2023currents,Reehorst:2024frustrated}
and supersymmetry \cite{Atanasov:2022bpi,Chang:2019dzt}. For ABJM theory,
precision islands for OPE coefficients
\cite{Chester:2014mea,Agmon:2017xes,Agmon:2019imm} have yielded a prediction
for the $D^8R^4$ interaction, the first unprotected higher derivative
correction to the M theory effective action \cite{Chester:2024mtheory}.
In four dimensions, numerical bootstrap studies of
$\mathcal N=4$ super Yang Mills theory have progressed from single and mixed
correlators to constraints across the conformal manifold
\cite{Beem:2013qxa,Beem:2016wfs,Bissi:2020jve,Chester:2021aun}.
More recently, crossing symmetry has been combined with integrability and
localization to obtain precise results across the coupling range
\cite{Caron-Huot:2022sdy,Chester:2023ehi,CaronHuot:2024localization}.

Numerical bounds are obtained by acting on crossing equations with a finite
space of linear functionals subject to suitable positivity conditions.
Their efficiency depends not only on the number of constraints, but also on
the linear functionals used to impose them. The standard choice is a basis of
 derivatives with respect to cross-ratios $(z,\bar z)$ at the crossing symmetric point,
$z=\bar z=1/2$ \cite{Rattazzi:2008pe,Poland:2022qrs,Rychkov:2023wsd,Rychkov:2025boot}. This
basis is simple and universal, but it probes crossing locally. The crossing
vectors are analytic functions on a cut plane, and their singularities and
discontinuities are accessed only indirectly by a Taylor expansion around one
point. At high derivative order, the local data must in effect reconstruct
this global analytic structure. The resulting bounds converge particularly
slowly when the external scaling dimension is large.

This limitation becomes important well before external dimensions are
parametrically large. Gauge invariant and monopole operators in conformal
gauge theories and deconfined critical points can already have scaling
dimensions substantially larger than those of familiar low lying Ising
operators \cite{He:2023currents,Chester:2025uxb}. A direct large external
dimension limit appears for a massive QFT in AdS. As the curvature radius
grows, dimensions of boundary operators scale with it and their correlators
recover flat space scattering \cite{Paulos:2016fap}.
In higher dimensions, the scalar unitarity bound itself raises the minimum
external dimension. Studies of $O(N)$ models in $4<d<6$ and of the six
dimensional $(2,0)$ theory illustrate this regime, while maximally
supersymmetric bootstrap calculations in four and six dimensions have
required very high derivative order for convergence
\cite{Chester:2014gqa,Beem:2015aoa,Alday:2022ldo}.

Dispersive analytic functionals provide alternatives to the derivative
basis
\cite{Gopakumar:2016cpb,Gopakumar:2016wkt,Mazac:2019shk,Sleight:2019ive, Penedones:2019tng,Carmi:2020ekr,Caron-Huot:2020adz,Gopakumar:2021dvg}.
Their adaptation to the numerical gap maximization problems considered here
is less direct because their actions do not generally furnish positivity
conditions essential for numerical optimization. %\red{Product analytic functionals were introduced in two dimensions in Ref.~\cite{Mazac:2016qev}, generalized and
%studied in higher even dimensions in Refs.~\cite{Paulos:2019gtx,Kaviraj:2021cvq}, and
%first proposed for odd dimensions in our previous work~\cite{Ghosh:2023onl}}
The product functionals were first explored in two dimensions in \cite{Mazac:2016qev}, yielding rigorous twist bounds. The basis underlying our work was introduced in~\cite{Paulos:2019gtx} in $d=2$, and gave preliminary evidence of its advantage over the derivative basis in some regime. 
We proposed the dimensional-reduction route to extend this program to higher dimensions in \cite{Ghosh:2023onl}. In this paper, we overcome the technical challenges that previously limited the
construction and complete its implementation for scalar four-point functions
in three dimensions. The
product analytic functionals are defined through contour integrals around the
branch cuts of the crossing vector and directly probe conformal block
discontinuities. They provide a higher-dimensional counterpart of the analytic
functionals that have proved very effective in one-dimensional bootstrap problems
\cite{Mazac:2016qev,Mazac:2018mdx,Mazac:2018ycv,Paulos:2019fkw,Ghosh:2025sic}.

The key step is to evaluate functional action on three-dimensional conformal blocks,
which are not available in closed form. Dimensional reduction
\cite{Hogervorst:2016hal} expresses each block as an infinite sum of two-dimensional blocks, and the functional action on every term factorizes into
one-dimensional functionals. We analytically resum the slowly converging tail. %at large level. 
%We also verify the convergence and swappability conditions that
%justify exchanging the functional action with the OPE sum. 
Although we focus
on three dimensions, the same strategy can be extended to other
spacetime dimensions.

Numerically, the functional actions generated above are interpolated and then
used as input to the bootstrap optimization. We solve the resulting semi infinite linear
program with an adaptive mixed-precision simplex method inside an outer
approximation loop. The loop solves a sequence of finite linear programs and
inserts new operators wherever positivity is violated. The software,
interpolation, refinement procedure, and production parameters are described
in \smref{app:analytic-numerics}.

We benchmark the product functional basis in scalar and spin-two gap
maximization. At the same basis size, it always produces stronger bounds and
converges more rapidly than the derivative basis, with the advantage becoming
more pronounced at larger external dimension. Already at the Ising point, where the two bases
are closest in performance, the derivative basis determination of the
$\Delta_\epsilon$ bound at $\Lambda=43$ ($253$ components) takes about ten
hours on an HPC node, whereas the analytic basis reaches the same accuracy at
$\lambda=5$ ($72$ components) in about thirty seconds on a laptop. The
improved scalar bound strengthens the
Nakayama and Ohtsuki necessary condition \cite{Nakayama:2016jhq} for a critical point with one tuning
parameter to $\Delta_0>1.06163$. In the spin-two problem, the analytic basis
resolves a new sharp kink near $\Delta_\phi\simeq4.16$, accompanied by a
reorganization of the extremal spectrum. Both bounds also display plateau
structures at larger external dimension that have not stabilized in the
available derivative calculations.

We use the extremal spectra to test whether distinguished points on the
boundary of the allowed region %boundary solutions 
coincide with familiar physical theories. At the Ising external dimension,
independent extrapolations in the analytic and derivative bases converge to a
scalar gap slightly above the Ising value. The %boundary 
extremal spectrum reproduces
the leading Ising data but misses or displaces higher operators.
At $\Delta_\phi=2$, the scalar %boundary 
extremal spectrum likewise does not coincide
with the free Majorana theory, although the two are quite close. %In the spin-two problem, mean-field-theory organization begins to deform before the new kink, and no CFT interpretation of the branch beyond it is presently known. 
 These examples demonstrate that
proximity to, or apparent saturation of, a numerical bound does not by itself
imply exact realization by a physical CFT. Rather, the corresponding extremal
solution may provide a close approximation that can be systematically refined
by enlarging the set of crossing equations.

\section{Product functionals}
\label{sec:prodfunc}

For identical external scalars with dimension $\Delta_\phi$, the crossing equation takes the following form,
\begin{equation}
  \sum_{\mathcal O}
  f_{\phi\phi\mathcal O}^{\,2}
  F^{d}_{\Delta,\ell}(z,\bar z)=0,\quad f_{\phi\phi\mathcal O}^{\,2}\geq 0
  \label{eq:crossing-3d}
\end{equation}
where \(F^{d}_{\Delta,\ell}(z,\bar z)\) is the crossing vector associated
with a primary of dimension \(\Delta\) and spin \(\ell\), and $f_{\phi\phi\mathcal O}^{\,2}$ is square of OPE coefficients.The dimensions, spins, and OPE coefficients of the operators appearing in the $\phi\times\phi$ OPE constitute the CFT data.
%Numerically, a finite set of linear functionals projects Eq.~\eqref{eq:crossing-3d} to finitely many scalar equations.
To do numerical conformal bootstrap, we act with a finite set of linear functionals on Eq.~\eqref{eq:crossing-3d}, which yields finitely many constraint equations. We can constrain CFT data by optimizing linear functionals subject to positivity on the assumed spectrum. The rate of convergence therefore depends on the functional space used.
% Bounds follow by optimizing their linear combinations subject to positivity on the assumed spectrum, so convergence depends on the chosen family.  
%In this paper, we use product analytic functionals, which construct higher dimensional actions from product of pairs of one dimensional functionals actions.  They are the only known analytic functional actions  in \(d>1\) for which strong evidence supports all five practical criteria for numerical applications of Ref.~\cite{Ghosh:2023onl}: finiteness, swapping, completeness, asymptotic positivity, and efficient computability.
In this paper, we study product analytic functionals, whose higher-dimensional
actions are obtained by taking products of pairs of one-dimensional functional
actions. To our knowledge, they are the only analytic functionals in \(d>1\)
for which strong evidence supports all five practical criteria for numerical
applications described in Ref.~\cite{Ghosh:2023onl}: finiteness, swapping,
completeness, asymptotic positivity, and efficient computability.

For a function with a cut along \([1,\infty)\), we define the discontinuity,
\begin{equation}
  \mathcal I_z X(z)
  \equiv
  \lim_{\epsilon\to0^+}
  \frac{X(z+i\epsilon)-X(z-i\epsilon)}{2i}.
  \label{eq:cut-discontinuity}
\end{equation}
%A one dimensional functional of crossing parity \(\pm\) can then be written as
The one-dimensional functional is written as the following integral over the discontinuity across the cut \cite{Mazac:2018ycv, Paulos:2019gtx},
\begin{equation}
  \omega^\pm_m(\delta\mid\delta_\phi)
  =
  \int_1^\infty\frac{dz}{\pi}\,
  h^\pm_m(z)\,
  \mathcal I_z F^\pm_\delta(z\mid\delta_\phi),
  \label{eq:oned-cut-functional}
\end{equation}
where \(F^-_\delta\)(z) is crossing antisymmetric and \(F^+_\delta\)(z) is crossing symmetric vector. The kernels $h^\pm_m(z)$ are constructed so that the required duality properties hold as explained in Appendix \ref{app:oned}, together with appropriate boundary falloff conditions at the endpoints of the integration contour \cite{Mazac:2018ycv}. When these functionals are used below, the one-dimensional external dimension is $\delta_\phi=\Delta_\phi/2$.  %\(\delta_\phi=\Delta_\phi/2\).  
In this work we use the bosonic family of functionals for the \(``-"\) type and the fermionic family for the \(``+"\) type. Within each family, the non-negative integer \(m\) labels the individual functionals. Further details are given in Appendix~\ref{app:oned}.

%These constituents 
Using these $1\text{D}$ functionals, we can define a product functional that acts on a crossing vector in any spacetime dimension
\(d\geq2\) \cite{Paulos:2019gtx,Ghosh:2023onl}:
\begin{small}
\begin{equation}
  (\omega^-_m\otimes\omega^+_n)
  \left[F^d_{\Delta,\ell}(z,\bar z\mid\Delta_\phi)\right]
  \equiv
  2\int_{++}\frac{dz\,d\bar z}{\pi^2}\,
  h^-_m(z)h^+_n(\bar z)
  \bigl[
    \mathcal I_z\mathcal I_{\bar z}
    F^d_{\Delta,\ell}(z,\bar z\mid\Delta_\phi)
    +
    \mathcal I_z\mathcal I_{\bar z}
    F^d_{\Delta,\ell}(z,1-\bar z\mid\Delta_\phi)
  \bigr].
  \label{eq:product-cut-definition}
\end{equation}
\end{small}
Both integrations run along $[1,\infty)$. Note that, without loss of generality, we symmetrize the crossing vector under $\bar z\leftrightarrow 1-\bar z$ before acting on it with the product of one-dimensional functionals. %The second term projects under \(\bar z\mapsto1-\bar z\). 
In two dimensions, factorization of conformal block reduces the double integral to the following product of $1\text{D}$ functional actions,
\begin{equation}
  (\omega^-_m\otimes\omega^+_n)
  \left[
    F^{d=2}_{\Delta,\ell}(z,\bar z\mid\Delta_\phi)
  \right]
  =
  \frac12
  \left[
    \omega^-_m\left(
      \frac{\Delta-\ell}{2}\mathrel{\bigg|}\frac{\Delta_\phi}{2}
    \right)
    \omega^+_n\left(
      \frac{\Delta+\ell}{2}\mathrel{\bigg|}\frac{\Delta_\phi}{2}
    \right)
    +(\ell\to-\ell)
  \right].
  \label{eq:product-factorization}
\end{equation}

For %three-dimensional 
3D conformal blocks, dimensional reduction
\cite{Hogervorst:2016hal} gives
\begin{equation}
  G^{(3d)}_{\Delta,\ell}
  =
  \sum_{q=0}^{\infty}
  \sum_{s=\ell,\ell-2,\ldots,\ell\bmod 2}
  A_{q,s}(\Delta,\ell)\,
  G^{(2d)}_{\Delta+2q,s}.
  \label{eq:dimensional-reduction}
\end{equation}
 Therefore, the functional action on a $3\text{D}$ crossing vector can be evaluated as infinite sum of $2\text{D}$ product functional actions,
\begin{small}
\begin{equation}
  (\omega^-_m\otimes\omega^+_n)
  \left[
    F^{d=3}_{\Delta,\ell}(z,\bar z\mid\Delta_\phi)
  \right]
  =
  \sum_{q,s}A_{q,s}(\Delta,\ell)\,
  (\omega^-_m\otimes\omega^+_n)
  \left[
    F^{d=2}_{\Delta+2q,s}(z,\bar z\mid\Delta_\phi)
  \right].
  \label{eq:3d-product-action}
\end{equation}
\end{small}

The sum over $q$ converges only as a power law and is therefore numerically challenging. We accelerate the convergence by subtracting the large-$q$ expansion of the summand, whose asymptotic tail can then be summed analytically in terms of Hurwitz zeta functions. In our numerical application we take $\lambda\in\mathbb{Z}_{\geq 0}$ and adopt the bosonic--fermionic product functional of \cite{Ghosh:2023onl}, with $\omega^-\in\{\alpha_B^-,\beta_B^-\}$ and $\omega^+\in\{\alpha_F^+,\beta_F^+\}$, retaining the functional pairs with $m+n\leq\lambda$. This yields $2(\lambda+1)^2$ number of functionals. Further implementation details are given in \smrefs{\ref{app:product}, \ref{app:asympt}, and \ref{app:analytic-numerics}}.

\section{Landscape of gap bounds}

% Figure 1 lives in sections/fig-gap-bounds.tex; the wrappers decide where
% it enters the source stream (here for arXiv, after \maketitle for PRL).
% Figure 1 (gap-bounds landscape).  Kept in its own file so each wrapper
% can choose the input position: the PRL letter inputs it right after
% \maketitle so the float can reach page 2, while the arXiv version inputs
% it at the top of the Landscape section via \placegapboundsfigure.
\begin{figurewidetop}
  \centering
  \begin{subfigure}[t]{0.49\linewidth}
    \centering
    \includegraphics[width=\linewidth]{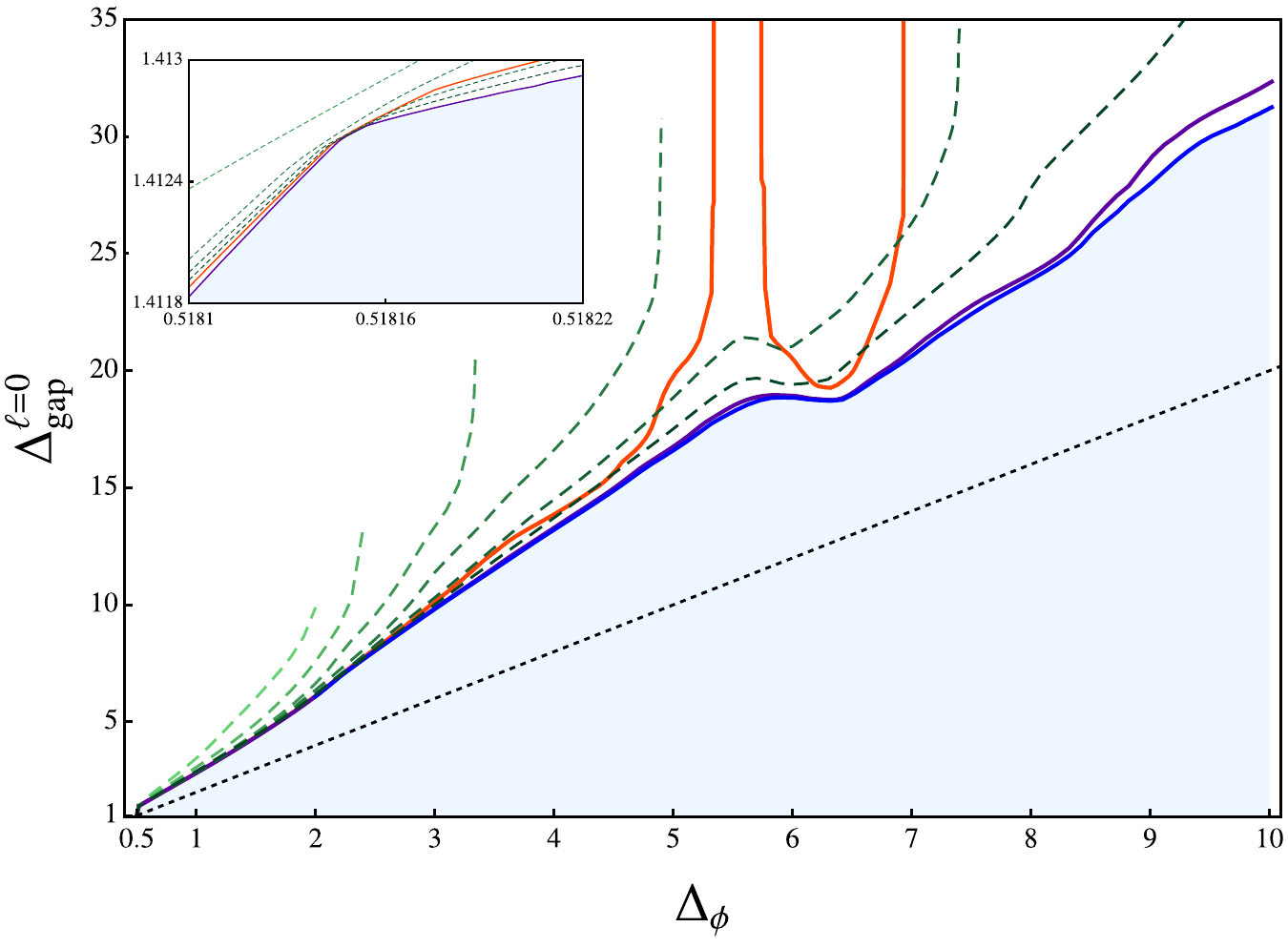}
    \caption{}
    \label{fig:scalar-bound}
  \end{subfigure}
  \hfill
  \begin{subfigure}[t]{0.49\linewidth}
    \centering
    \includegraphics[width=\linewidth]{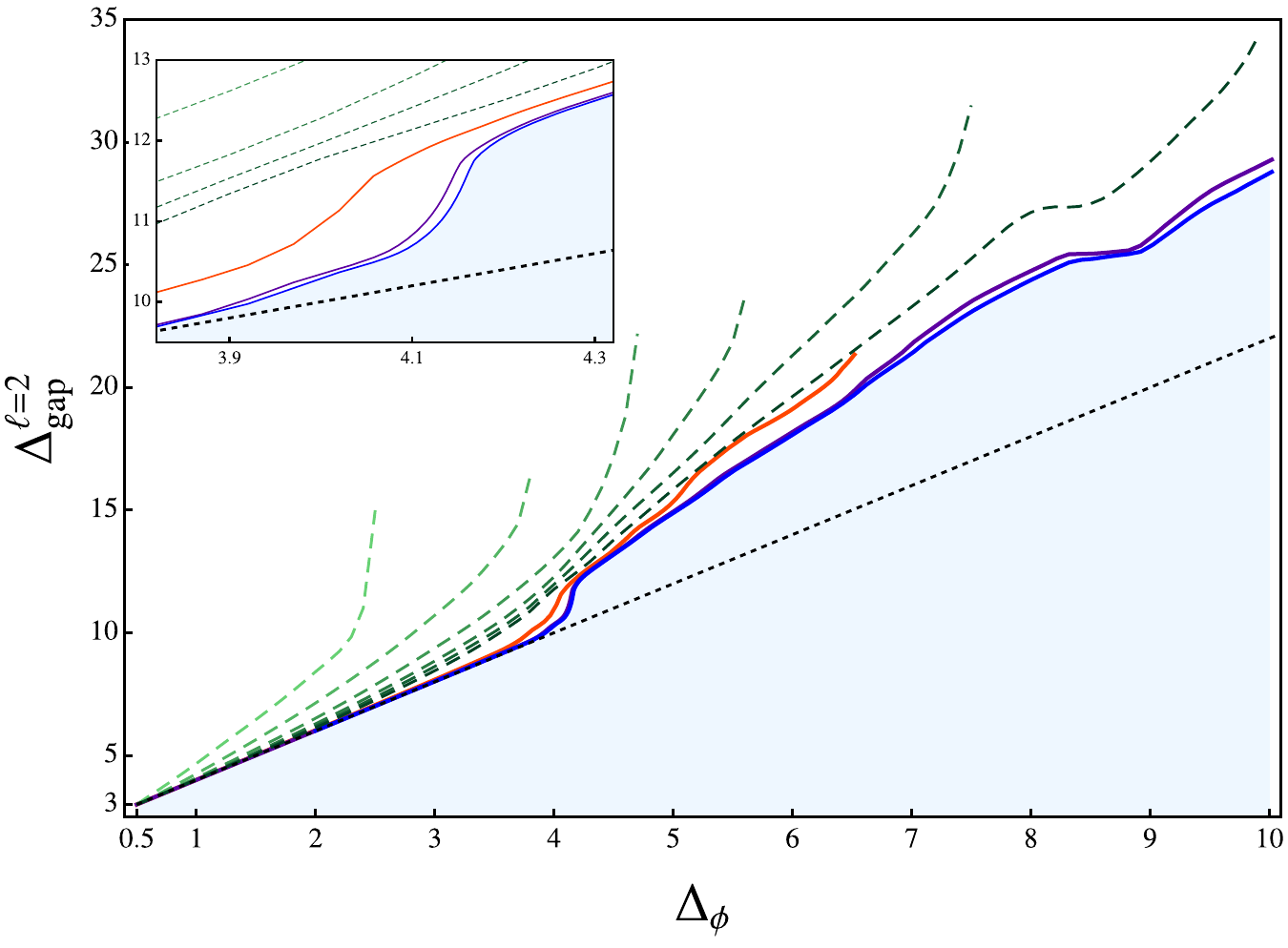}
    \caption{}
    \label{fig:spin2-bound}
  \end{subfigure}
  \caption{%
    Upper bounds on gaps in the $\phi\times\phi$ OPE in $d=3$ as
    functions of the external dimension $\Delta_\phi$.  Panel (a) shows the
    leading scalar gap, while panel (b) shows the gap for operators of spin
    two without imposing an isolated stress tensor.  Solid curves show the
    analytic functional results at $\lambda=5,10,12$, corresponding
    respectively to basis sizes $N=72,242,338$, and are ordered from warm
    orange to cool blue.
    Dashed green curves show the derivative basis results at
    $\Lambda\in\{7,11,19,29,43,63\}$, corresponding respectively to
    $N\in\{10,21,55,120,253,528\}$, and are ordered from light to dark.
    A complete comparison including every analytic functional cutoff from
    $\lambda=1$ to $12$ is shown in \smfigref{fig:gap-benchmark}.
    The black dotted line marks the leading generalized-free-field double-twist dimension, $2\Delta_\phi$ in panel (a) and $2\Delta_\phi+2$ in
    panel (b).  The light blue region below the bound curves is allowed.
    Panel (a) marks the three-dimensional Ising point.  Its inset resolves it
    near
    $(\Delta_\phi,\Delta_\epsilon)\approx(0.518149,1.412626)$.  The inset in
    panel (b) resolves the kink region
    $\Delta_\phi\in[3.82,4.32]$.%
  }
  \label{fig:gap-bounds}
\end{figurewidetop}

Now we compare the product analytic functional basis with the standard derivative basis in single-correlator gap-maximization problems for identical external scalars $\phi$ in three-dimensional CFT, over the range
$\Delta_\phi\in[0.5,10]$. We consider the scalar-gap and spin-2 gap maximization problems separately, without imposing an isolated stress
tensor. We use up to $338$ components $(\lambda=12)$ in the analytic-functional basis and up to \(528\) components $(\Lambda=63)$ in the derivative-basis  using \texttt{SDPB}. The two numerical implementations are described in \smref{app:analytic-numerics} and \smref{app:sdpb-parameters}, respectively.
Figure~\ref{fig:gap-bounds} shows the scalar and spin-two upper bounds as functions of $\Delta_\phi$. As we increase the cutoff, the bounds obtained from both bases decrease, but at comparable basis size, the bounds from the product-functional decrease more rapidly, and the difference between bounds obtained from the two bases grows with $\Delta_\phi$.
At small $\Delta_\phi$, the two bases give nearly identical bounds in both
the scalar- and spin-two-gap maximization problems. In the scalar case, they
reproduce the well-known three-dimensional Ising kink, shown in greater
detail in the inset of Fig.~\ref{fig:scalar-bound}. We analyze the extremal
spectrum at the Ising point in \smsecref{sec:ising-kink}. In the spin-two
channel, the sharper bound obtained with the product-functional reveals
a kink near $\Delta_\phi\approx4.16$, which, to our knowledge, has not been
observed before. The kink is already visible at modest cutoff, as shown in
the inset of Fig.~\ref{fig:spin2-bound}. The full extremal spectrum shows that GFF saturates it at smaller $\Delta_\phi$ and begins to depart from it for $\Delta_\phi\gtrsim3$. The limiting bound may therefore depart smoothly from the GFF value before reaching the kink. The feature near $4.16$ marks a later,
sharp reorganization of the extremal spectrum, which we analyze in \smsecref{sec:gff}.

At large $\Delta_\phi$, finite-cutoff behavior of the two bases becomes visibly different.  In panel (a), the
$\lambda=5$ curve develops spurious spikes for $\Delta_\phi$ around $5.5$.  These truncation artifacts are absent at $\lambda=10,12$.  At the two largest basis sizes in the analytic functional basis, both channels display plateau structure: the scalar bound stalls near $\Delta_\phi\approx6.5$, where it is mildly nonmonotonic, and the spin-two bound stalls near $\Delta_\phi\approx8.5$, before resuming its growth.  The persistence of these features from $\lambda=10$ to $12$ provides evidence that they survive in the limiting bounds, although we do not attempt an extrapolation in the cutoff. The bounds derived using the derivative basis have not stabilized in this region yet.

The sharper scalar bound has an immediate physical application.  A critical
point that can be reached by tuning a single parameter must contain no
relevant singlet scalar other than the leading one, $\mathcal{O}_0$: the next
singlet scalar in the $\mathcal{O}_0\times\mathcal{O}_0$ OPE must be
irrelevant, i.e.\ $\Delta^{\ell=0}_{\mathrm{gap}}\geq 3$ at
$\Delta_\phi=\Delta_0$.  The point where the gap bound of
Fig.~\ref{fig:scalar-bound} crosses $3$ therefore yields a necessary
condition on $\Delta_0$.\footnote{Strictly speaking, this bound corresponds to maximizing the gap to the operator above $\mathcal{O}_0$. Following \cite{Nakayama:2016jhq}, however, this bound coincides with the scalar channel gap maximization we studied in Figure~\ref{fig:gap-bounds}, as we can verify numerically: the condition $\alpha[F^{\Delta_0}_{\Delta_0,0}]\geq0$ is satisfied by the extremal functional obtained from gap maximization.}  From the derivative basis,
Nakayama and Ohtsuki obtained the well-known bound $\Delta_0>1.044$, equivalently critical exponent $\nu=1/(3-\Delta_0)>0.511$ \cite{Nakayama:2016jhq}.  At our highest cutoff, $\lambda=12$, in analytic functional basis the bound sharpens both numbers to
\begin{equation}
  \Delta_0>1.06163,
  \qquad
  \nu>0.51590.
  \label{eq:NO-bound}
\end{equation}
This tightens the constraint on critical points with one tuning parameter and on the emergent symmetry scenarios it was devised to test.

\section{Convergence}
% Figure 2 lives in sections/fig-spin2-convergence.tex; the wrappers decide
% where it enters the source stream (here for arXiv, earlier for PRL).
% Figure 2 (spin-2 convergence).  Kept in its own file so each wrapper can
% choose the input position via \placespinconvergencefigure (both versions
% currently anchor it at the top of the Convergence section).  The
% figuremain environment and \convergencefigwidth are defined per wrapper.
% NOTE: do not switch figuremain to an explicit [t] in the PRL wrapper --
% revtex4-2 under-measures the caption of explicit top floats and the
% caption overlaps the following text; [h] placement measures correctly.
\begin{figuremain}
  \centering
  \includegraphics[width=\convergencefigwidth]{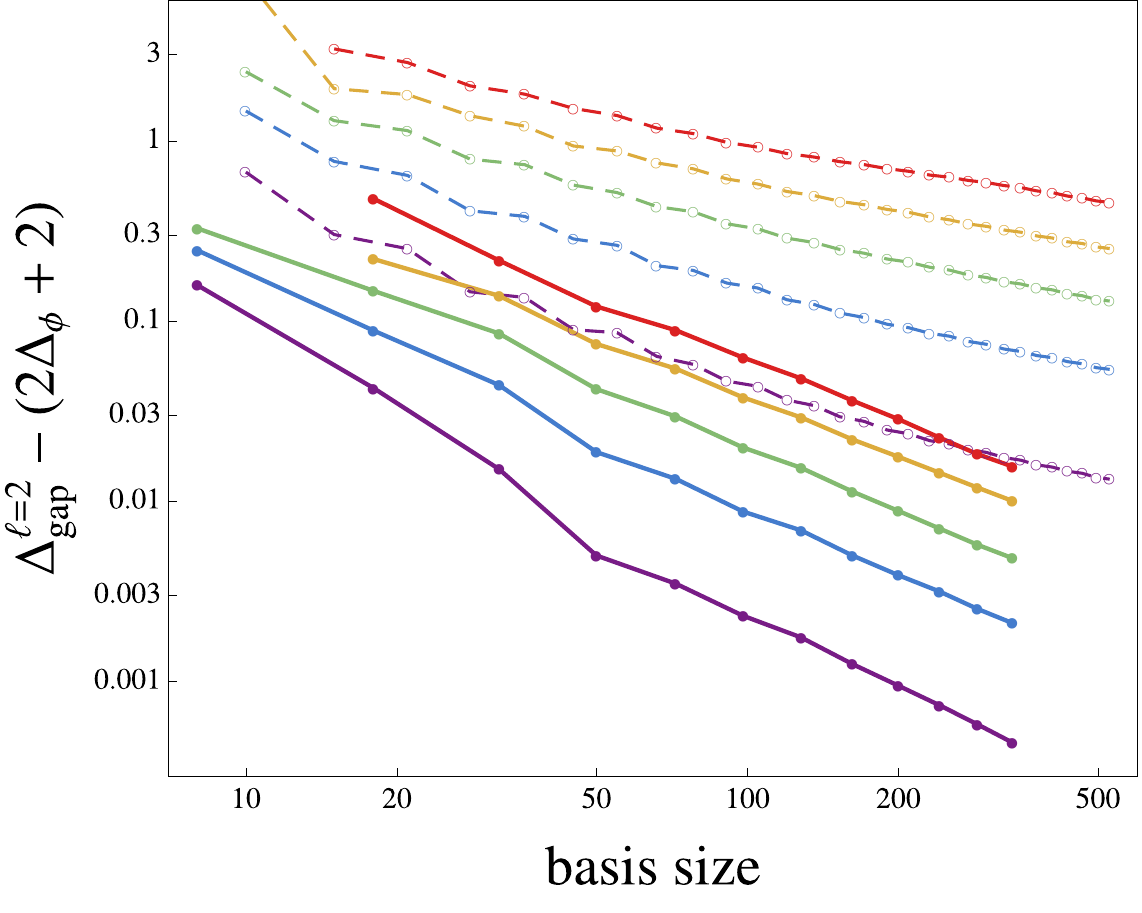}
  \caption{%
    Residual of
    the leading gap for operators of spin two relative to the leading GFF
    double-twist value, plotted against the basis size $N$ on logarithmic
    axes.  Solid curves with filled markers show the analytic functional
    basis, and dashed curves with open markers the derivative basis.  Colors
    denote $\Delta_\phi=1,\tfrac32,2,\tfrac52,3$, ordered from purple to
    red.%
  }
  \label{fig:spin2-convergence}
\end{figuremain}

We compare the convergence rates in Fig.~\ref{fig:spin2-convergence}.  The
spin-2 gap is well suited to this test because the conjectured optimal
value $2\Delta_\phi+2$ (see \smsecref{sec:gff}) provides a sharp reference
against which residuals can be measured. For
$\Delta_\phi=1,\tfrac32,2,\tfrac52,3$, the residual
$\bigl|\Delta^{\ell=2}_{\mathrm{gap}}-(2\Delta_\phi+2)\bigr|$ decreases
approximately as a power of the basis size for both methods.  The two
sequences are consistent with convergence to the same generalized-free-field
value.  The analytic functional basis converges more rapidly: with a few
hundred components, its residual is one to two orders of magnitude smaller
than that obtained with the derivative basis.  Writing the residual as
$N^{-p}$, the effective slopes over the displayed range give approximately
$p=1.4$ to $1.1$ for the analytic functional basis and $p=0.7$ to $0.4$ for
the derivative basis.

One possible explanation lies in how the two bases probe the heavy spectrum.
At the crossing-symmetric point, individual derivatives of a conformal block
are exponentially suppressed at large exchanged dimension, up to polynomial
prefactors, as expected from exponential convergence of the OPE
\cite{Pappadopulo:2012jk}.  By contrast, the large-dimension actions of
the analytic functional basis have polynomial asymptotics, as reviewed in
\smref{app:asympt}.  The resulting sensitivity to heavy operators provides a
plausible interpretation of the steeper slopes in
Fig.~\ref{fig:spin2-convergence}, particularly at large $\Delta_\phi$, where
the double-twist operators controlling the bound are themselves heavy.

A similar advantage appears in the extremal spectra studied in
\smsecref{sec:extremal}.  The analytic basis reaches stable estimates for
subleading dimensions at smaller basis size, and the improvement becomes more
pronounced for heavier operators.  The gain is therefore not confined to the
leading operator in the extremal spectrum. %gap.

\section{Extremal solutions versus physical theories}
\label{sec:extremal}

The boundary of a gap maximization problem is characterized by an
\emph{extremal solution} to crossing. Determining whether such a solution
coincides with a familiar physical theory has attracted considerable
interest in the bootstrap community. The improved convergence of the
analytic functional basis, both in the leading and subleading operators, allows us to test this issue using both the optimized gap and
the associated extremal spectrum. We consider three benchmark comparisons.
In the scalar gap problem, we compare the extremal solutions at the Ising
kink and at \(\Delta_\phi=2\) with, respectively, the three-dimensional Ising
CFT and the free Majorana theory, where the external scalar is the mass
operator. We then compare the boundary of the gap problem in the sector of
spin two with the mean-field-theory solution. The extremal spectra at the Ising kink and at $\Delta_\phi=2$ closely resemble those of the corresponding reference theories, but a closer inspection reveals clear differences.
%exhibit resolved structural differences.
By contrast, the results in the
sector of spin two provide strong evidence for mean-field-theory saturation
over a wide range where stable double-twist
operators appear in the extremal spectrum. %organization remains stable. 
Together,
these examples show that close agreement in the leading data is not sufficient to
identify an extremal solution with a known theory.

\subsection{The Ising CFT and the scalar gap boundary}
\label{sec:ising-kink}

The proximity of the three-dimensional Ising CFT to the kink in the scalar
gap bound has long suggested that the Ising CFT might exactly saturate the
optimal bound~\cite{El-Showk:2012cjh,El-Showk:2014dwa}.  The improved
convergence obtained here allows us to test this conjecture at substantially
higher precision.  We find that the Ising CFT lies remarkably close to the
single-correlator bound, but does not saturate it.  
 Two complementary
observations lead to this conclusion.  First, independent extrapolations in
the analytic and derivative bases converge to a scalar gap slightly above the
Ising value. Second, the corresponding extremal spectra agree for the low-lying operators, but some operators present in the Ising CFT are absent from the extremal spectrum.  %Second, the corresponding extremal spectra agree for the leading operators but differ structurally from the Ising spectrum at higher dimension.

\begin{figure}[t]
  \centering
  \includegraphics[width=0.82\linewidth]{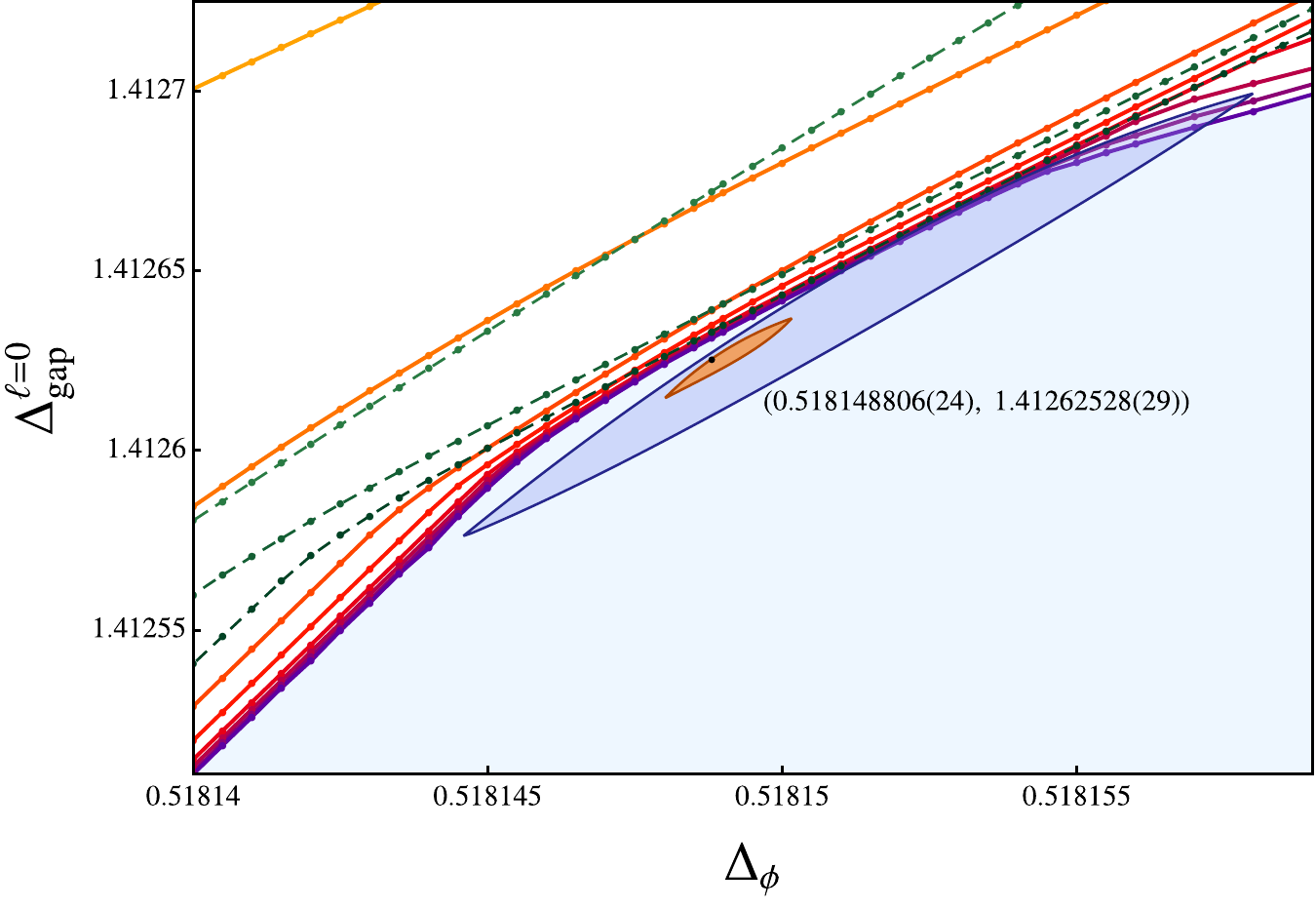}
  \caption{%
    Ultra-zoom of the spin-0 ($\ell=0$) gap bound around the $3$d Ising point,
    $\Delta_\phi\in[0.518140,\,0.518159]$.
    The \emph{analytic-functional} cutoffs $\lambda$ up to $10$ (solid, orange$\,\to\,$red$\,\to\,$violet
    with increasing $\lambda$) and the \emph{derivative-basis} truncations
    $\Lambda=2n_{\max}-1\in\{29,43,63\}$ (dashed green, light to dark with
    increasing $\Lambda$) are as in the full benchmark of \smref{app:funconv}; the
    markers indicate
    the actual computed data points, and the light-blue region below every curve is
    allowed.
    The broad blue and smaller orange islands are obtained from the
    \(\sigma,\epsilon\) mixed-correlator bootstrap at \(\Lambda=43\)
    \cite{Simmons-Duffin:2015qma,Kos:2016ysd}.  The smaller island additionally
    scans the ratio
    \(f_{\epsilon\epsilon\epsilon}/f_{\sigma\sigma\epsilon}\),
    thereby imposing uniqueness of the relevant operators.  The black dot
    marks the rigorous determination from the enlarged
    \(\sigma,\epsilon,T\) mixed-correlator bootstrap
    \cite{Chang:2024whx},
    $(\Delta_\sigma,\Delta_\epsilon)=(0.518148806(24),\,1.41262528(29))$, toward which
    the single-correlator bounds descend from above.%
  }
  \label{fig:ising-benchmark}
\end{figure}

%\paragraph{A resolved separation from the bound.}
We fix the external dimension to
\(\Delta_\phi=0.5181488023\), which lies
within the current best determination of the scaling dimension of the spin operator in the 3D Ising CFT
\(\Delta_\sigma=0.518148806(24)\) \cite{Chang:2024whx}, and maximize the dimension of the first non-identity scalar appearing in \(\phi\times\phi\).  Figure~\ref{fig:ising-benchmark} shows that the
analytic functional and derivative basis bounds approach one another as their
respective cutoffs are increased.  For comparison, the figure also shows two
earlier mixed-correlator determinations using \(\sigma\) and \(\epsilon\) at
\(\Lambda=43\).  The broad blue island is the allowed region obtained in the original SDPB precision study \cite{Simmons-Duffin:2015qma}.  The smaller
orange island is the subsequent precision region of Ref.~\cite{Kos:2016ysd},
which scans the ratio
\(f_{\epsilon\epsilon\epsilon}/f_{\sigma\sigma\epsilon}\) and
thereby imposes uniqueness of the relevant operators.  The black point is
the later determination from an enlarged mixed-correlator system involving
\(\sigma\), \(\epsilon\), and the stress tensor \(T\)
\cite{Chang:2024whx}. Extrapolations of our two single-correlator
sequences give compatible asymptotic values near $\Delta_\epsilon^{\mathrm{Ising}}
  =1.41262528(29)~\cite{Chang:2024whx}$:
\begin{equation}
  \Delta_{\mathrm{gap},\infty}^{(0)}
  \simeq 1.412631(1).
  \label{eq:ising-gap-separation}
\end{equation}
Thus the %limiting 
extrapolated single-correlator bound remains above the Ising value by approximately
\(6\times10^{-6}\).  This separation is resolved independently by both
functional bases.   % The limiting value in Eq.~\eqref{eq:ising-gap-separation} should therefore be interpreted as the gap of a nearby extremal solution to crossing, rather than as another determination of \(\Delta_\epsilon\) in the Ising CFT.
The leading OPE data display the same remarkable proximity.  As shown in
Table~\ref{tab:ising_l0} of \smref{app:funconv}, the analytic extremal solution at \(\lambda=10\)
gives \(f_{\phi\phi\epsilon}^{\,2}=1.1063903\), compared with the Ising value
\(f_{\sigma\sigma\epsilon}^{\,2}\simeq1.1063962\).  The agreement of the
leading dimension and OPE coefficient explains why the Ising CFT appears to
saturate the bound at ordinary numerical precision.  The resolved gap
difference nevertheless shows that the two solutions are not identical.

\begin{tablewide}
  \centering
  \begin{tabular}{l c c c c}
    \toprule
    operator & $\ell$ & $\Delta_\infty$ (single corr.) & $\Delta$ (3d Ising) & $\Delta_\infty-\Delta_{\text{Ising}}$ \\
    \midrule
    $\epsilon$    & 0 & $1.4126311$ & $1.412625$ & $+6\times10^{-6}$   \\
    $\epsilon'$   & 0 & $3.82979$   & $3.82968$  & $+1.1\times10^{-4}$ \\
    $\epsilon''$  & 0 & $6.9453$    & $6.8956$   & $+5.0\times10^{-2}$ \\
    $T'$          & 2 & $5.50993$   & $5.50915$  & $+7.8\times10^{-4}$ \\
    \bottomrule
  \end{tabular}
  \caption{Single-correlator extremal limits $\Delta_\infty$---the values the
  single-correlator extremal spectrum converges to as $\lambda\to\infty$---for the
  low-lying scalars $\epsilon,\epsilon',\epsilon''$ and the subleading spin-2 $T'$,
  with the mixed-correlator $3$d Ising values and their difference.  Agreement is at
  the $10^{-4}$ level for $\epsilon,\epsilon'$ and $\sim10^{-3}$ for $T'$, but
  $\epsilon''$ is displaced by $\sim5\times10^{-2}$; the next scalar $\epsilon'''$
  is not resolved by the single correlator and is therefore absent here.}
  \label{tab:delta_inf}
\end{tablewide}

%\paragraph{A distinct extremal spectrum.}
The extremal spectrum gives a second and qualitatively different test.  If the Ising
CFT exactly saturated the scalar gap bound, the limiting extremal functional
would be expected to reproduce the operators in the
\(\sigma\times\sigma\) OPE.  Figure~\ref{fig:ising-spectrum} confirms this
expectation for the leading scalar \(\epsilon\), the stress tensor \(T\), the
next scalar \(\epsilon'\), the leading operator of spin \(4\), and the
subleading operator of spin \(2\). For all five operators, both functional bases approach the known Ising dimensions~\cite{Simmons-Duffin:2016wlq}. Our results are also consistent with those obtained using the fuzzy-sphere approach to the Ising spectrum, introduced in ~\cite{Zhu:2022gjc} and developed further in the recent study of finite-volume corrections and OPE data in~\cite{Lauchli:2025fii}. The agreement does not extend to the full spectrum. The scalar branch
identified with \(\epsilon''\) approaches a value displaced from the Ising
estimate, as quantified in Table~\ref{tab:delta_inf}. More decisively, no stable spin-zero operator can be identified with
$\epsilon'''$ at any of the available analytic or derivative cutoffs. Increasing the cutoff improves the convergence of the nearby operators, but none approaches the $\epsilon'''$ reference value. The agreement between the
two bases makes this a robust feature of the extremal spectrum.
%Since both bases show the same behavior, this is unlikely to result from a failure of operator tracking. 

\begin{figurewide}
  \centering
  \includegraphics[width=0.90\linewidth]{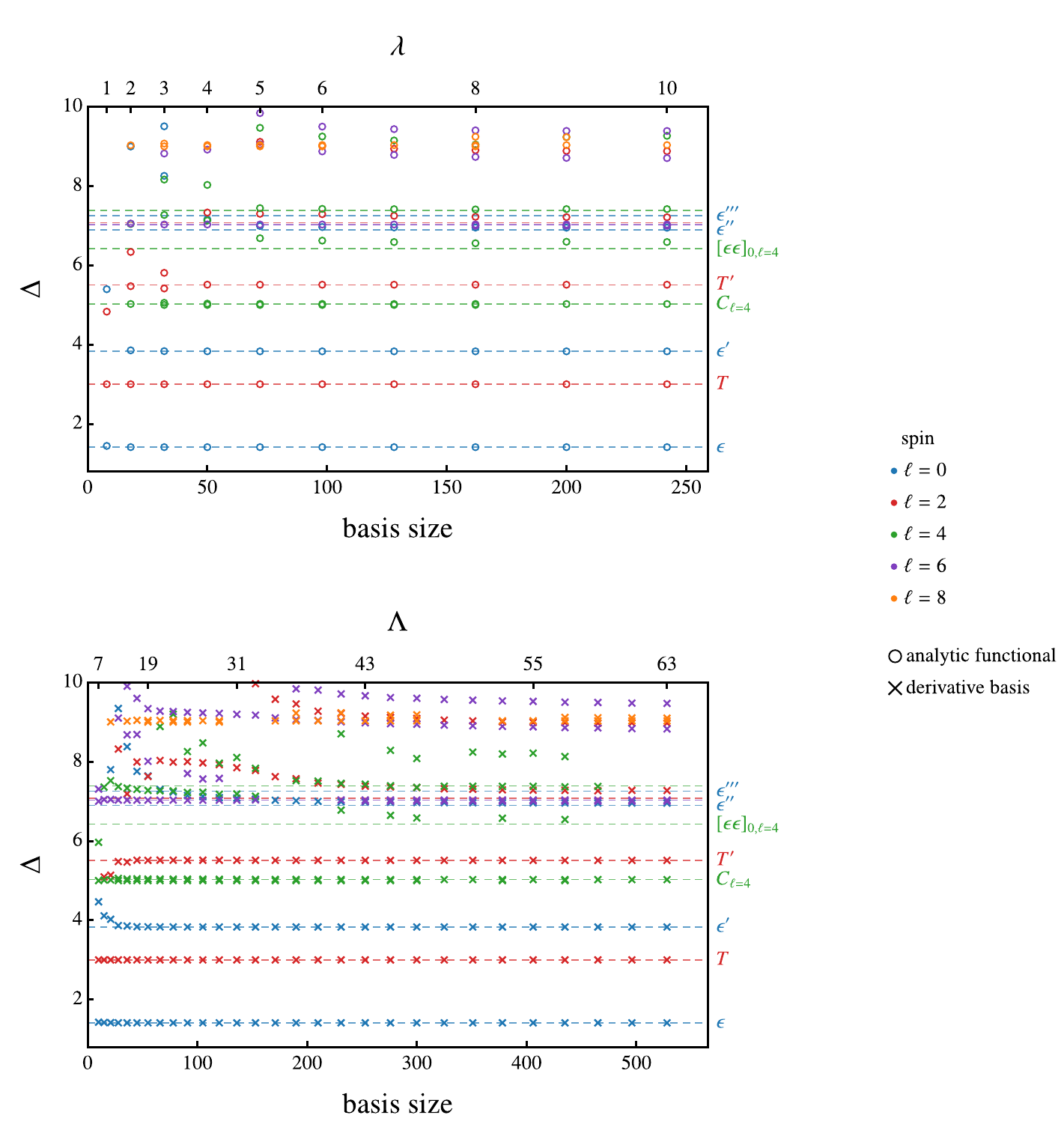}
  \caption{%
    Extremal spectrum extracted at the $3$d Ising point as a function of basis
    size.  The upper panel shows the analytic functional results as hollow circles,
    with the analytic cutoff $\lambda$ on the upper axis.  The lower panel shows
    the derivative basis results as crosses, with the derivative order $\Lambda$
    on the upper axis.  Colors denote spin, and dashed horizontal lines mark
    known Ising reference dimensions, including
    ($\epsilon,T,\epsilon',C_{\ell=4},T',[\epsilon\epsilon]_{0,\ell=4},
    \epsilon'',\epsilon'''$); the leading dimensions are from
    \cite{El-Showk:2014dwa}, the higher lines are shown as indicative reference
    values.  The operators of lowest dimension converge to their reference values
    in both bases.  At higher dimension, several levels converge to values
    displaced from the Ising references, while above $\Delta\approx7.5$ the
    extracted operators scatter and fail to stabilize.  In particular, no
    scalar branch converges to the $\epsilon'''$ line.%
  }
  \label{fig:ising-spectrum}
\end{figurewide}

\begin{figure}[t!]
  \centering
  \includegraphics[width=0.72\linewidth]{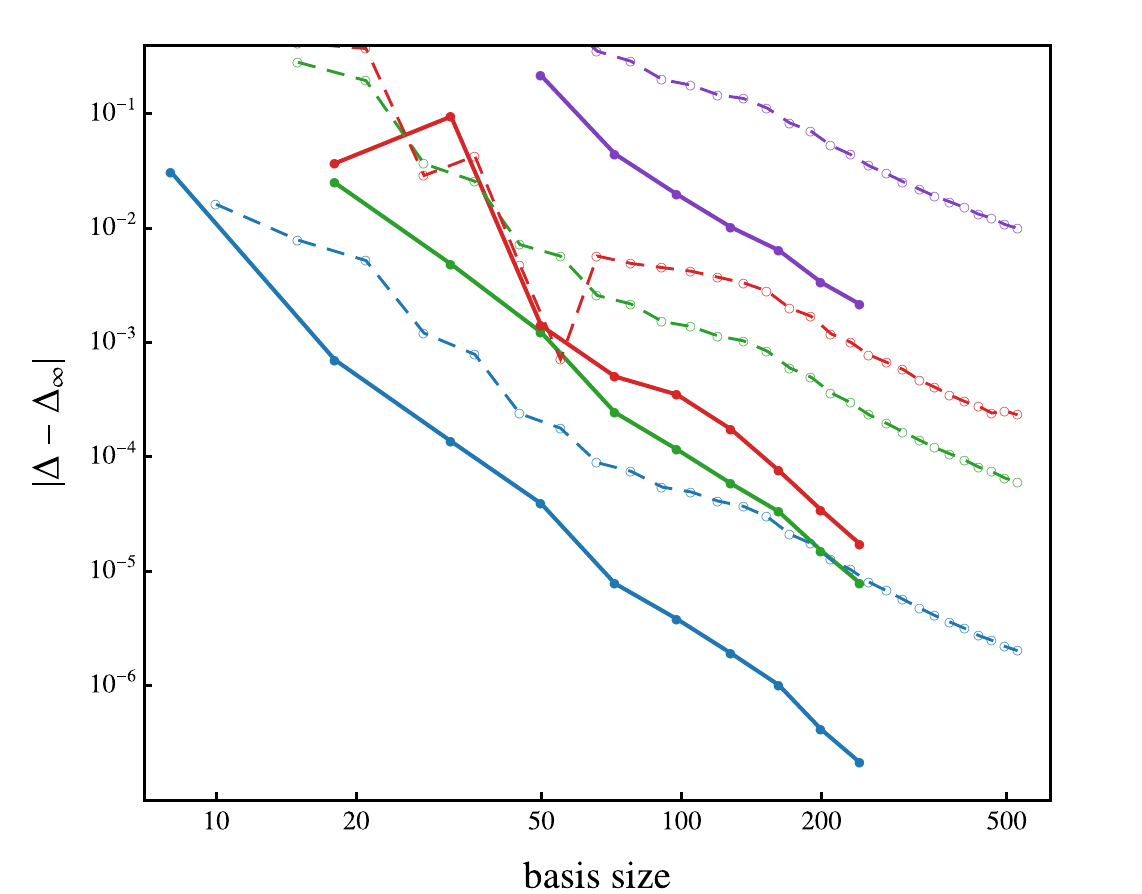}
  \caption{%
    Convergence of the extracted operator dimensions at the Ising external
    dimension toward their single-correlator limits $\Delta_\infty$.  The
    vertical axis shows $|\Delta-\Delta_\infty|$ and the horizontal axis
    shows basis size, both on logarithmic scales.  Colors distinguish
    $\epsilon,\epsilon',\epsilon''$, and $T'$.  Filled markers joined by
    solid curves show analytic functional data, while open markers joined
    by dashed curves show derivative basis data.  Both approaches exhibit
    power-law convergence to the same $\Delta_\infty$, while the analytic
    basis reaches the limit at smaller basis size, with the advantage
    increasing for heavier operators.  The limiting values are compared
    with the Ising spectrum in Table~\ref{tab:delta_inf}.%
  }
  \label{fig:ising-residual}
\end{figure}

%\paragraph{Convergence to distinct limits.}
%The disagreement with the higher Ising levels is not caused by a failure of numerical convergence.
The discrepancies at higher Ising levels persist as the cutoff is increased and are therefore not numerical artifacts. Figure~\ref{fig:ising-residual} shows the residual
$|\Delta-\Delta_\infty|$ relative to the fitted single-correlator limit for
each tracked operator.  The analytic functional and derivative basis
sequences approach the same limits with power-law behavior, while the
analytic basis reaches them at smaller basis size.  The gain increases for
the heavier operators.  These common limits do not all coincide with the
Ising data.  As summarized in Table~\ref{tab:delta_inf}, the discrepancies
for $\epsilon$, $\epsilon'$, and $T'$ are approximately
$6\times10^{-6}$, $1.1\times10^{-4}$, and $7.8\times10^{-4}$,
respectively, whereas the discrepancy for $\epsilon''$ is approximately
$5.0\times10^{-2}$.  Increasing the cutoff therefore sharpens convergence
to a neighboring extremal spectrum rather than restoring the full Ising
spectrum.

%\paragraph{Conclusion.}
The gap and spectrum tests probe different aspects of the extremal solution. The gap fits indicate that the Ising CFT lies slightly inside
the allowed region, although such extrapolations are not rigorous exclusions
by themselves.  The missing and displaced higher operators show
how the boundary correlator differs from the Ising correlator despite their
nearly identical leading data.  Taken together, these two independent
observations lead us to conclude that the three-dimensional Ising CFT does not exactly saturate the optimal single-correlator
scalar gap bound.  Mixed-correlator constraints isolate the physical Ising
solution, while the single-correlator extremal solution selects a neighboring
point on the boundary of the allowed space.

\paragraph{Sparseness of extremal solutions in ($d>2$):}
Our results also raise a broader question: how sparse can an extremal solution be? Here, by sparseness we mean the number of stable operator families that remain in the extremal spectrum at fixed spin as the functional cutoff is increased. In one dimension, extremal spectra arising in a wide class of gap-maximization problems are expected to approach the spectral and OPE density of generalized-free-field theory \cite{Ghosh:2025sic}. A similar statement holds in two-dimensional gap-maximization problems: although minimal models are special integrable solutions, their decomposition into global conformal blocks has the same asymptotic density of primary operators as 2D generalized-free-field theory.

The situation in higher dimensions is less clear. The usual Polyakov conditions constrain deformations of the generalized-free-field double-twist spectrum \cite{Caron-Huot:2020adz,Gopakumar:2021dvg}. However, higher-dimensional crossing also contains additional constraints associated with odd-spin functionals. These constraints cannot in general be satisfied by deforming only the ordinary double-twist families, and additional operator families are required to satisfy these sum rules\cite{KGpaper}. %This raises a natural question: do such additional families also appear as stable trajectories in higher-dimensional extremal solutions?
Our present numerical results give some evidence that additional stable trajectory appears in our extremal solution, see Table \ref{tab:multitwist-trajectory}. Besides the levels naturally associated with the double-twist families, we observe additional operators that persist as the functional cutoff is increased. %The clearest candidate appears near $\Delta\simeq 6.56$ at spin (4), with a possible continuation at spin (6). 
Their convergence is much slower, presumably because the corresponding OPE coefficients are strongly suppressed, so at present we do not claim a definitive identification of the full trajectory. Nevertheless, the fact that these levels do not disappear as the cutoff is increased suggests that they may be the first visible members of the additional families required by crossing of a single correlator. This question is closely related to the appearance of multitwist families required by higher-dimensional crossing at large spin \cite{Simmons-Duffin:2016wlq}. A more systematic study, following both their dimensions and OPE coefficients to larger cutoffs and spins, would be very interesting. 

\begin{figure}[h]
    \centering
    \includegraphics[width=0.78\textwidth]{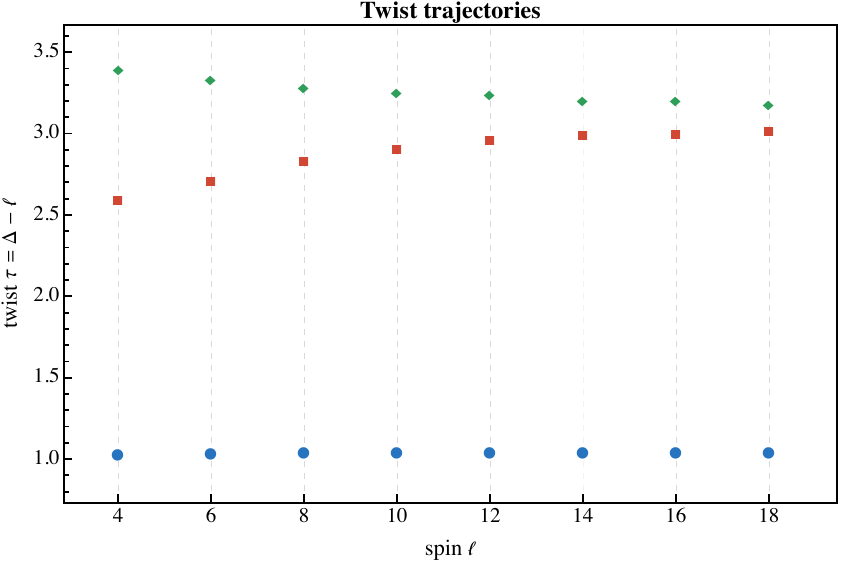}
    \caption{Twist trajectories $\tau=\Delta-\ell$ for three families of
    even-spin operators. Blue circles and green diamonds denote the leading
    and first subleading double-twist families,
    $[\sigma\sigma]_{0,\ell}$ and $[\sigma\sigma]_{1,\ell}$, respectively.
    Between them we find an additional tower, shown by red squares. The
    low-spin members of this intermediate tower have converged well, while
    the higher-spin dimensions are still converging as the basis size is
    increased. Within our present numerical resolution, however, we find no
    indication that the intermediate and upper trajectories merge.}
    \label{fig:three-operator-families}
\end{figure}

\begin{table}[t]
\centering
\begin{tabular}{c c c c}
\hline
Spin $\ell$ & $\Delta$ & Twist $\tau=\Delta-\ell$ & OPE coefficient \\
\hline
4  & 6.58602463  & 2.58602463 & $7.753959\times10^{-5}$ \\
6  & 8.70329198  & 2.70329198 & $6.119411\times10^{-5}$ \\
8  & 10.82460137 & 2.82460137 & $5.184116\times10^{-5}$ \\
10 & 12.90183241 & 2.90183241 & $4.432511\times10^{-5}$ \\
12 & 14.95267246 & 2.95267246 & $3.752212\times10^{-5}$ \\
14 & 16.98254688 & 2.98254688 & $3.124814\times10^{-5}$ \\
16 & 18.99409259 & 2.99409259 & $2.565642\times10^{-5}$ \\
18 & 21.00733935 & 3.00733935 & $2.149522\times10^{-5}$ \\
\hline
\end{tabular}
\caption{Candidate additional trajectory in the extremal spectrum beyond the double-twist trajectories. Nearly degenerate numerical roots have been combined by averaging their dimensions and summing their OPE coefficients.% The persistence of these operators from spin 4 to spin 18, with twist close to 2.82 ($(\epsilon \epsilon)_0$ family), provides evidence for an additional family beyond the double-twist trajectories. 
Their strongly suppressed OPE coefficients are consistent with the comparatively slow numerical stabilization.}
\label{tab:multitwist-trajectory}
\end{table}

\subsection{Comparison with the free Majorana theory at $\Delta_\phi=2$}
\label{sec:majorana}

A natural second benchmark for the scalar gap problem arises at
$\Delta_\phi=2$, where the reference theory is the free massless Majorana
fermion.  We identify the external scalar with the
mass operator
\begin{equation*}
  \epsilon=\psi^\alpha\psi_\alpha,
  \qquad
  \Delta_\epsilon=2.
\end{equation*}
The two-component nature of $\psi$ implies $\epsilon^2=0$.  A nonzero scalar
containing four fermion fields must instead contain at least two derivatives,
so its lowest possible dimension is six.  Such a scalar exists, together
with an operator of spin two at the same dimension.  The first non-identity scalar operator $\epsilon'$ in the OPE of $\epsilon$ has dimension $\Delta_{\epsilon'}=6$ 
%free Majorana scalar gap is therefore
rather than the generalized-free-field value $2\Delta_\epsilon=4$ \cite{Nakayama:2019jvm}. The free OPE has additional selection rules.  %Since $\epsilon$ is odd under time reversal, $\epsilon\times\epsilon$ contains only operators that are even under time reversal.  
Bose symmetry allows only %even spin 
operators with even spin in the OPE of $\epsilon\times\epsilon$. So Wick contractions
leave zero, two, or four uncontracted fermion fields.  The two-fermion sector
contains the even-spin conserved currents
\begin{equation*}
  J_\ell\sim\psi\gamma\partial^{\ell-1}\psi,
  \qquad
  \Delta_{J_\ell}=\ell+1,
  \qquad
  \ell=2,4,6,\ldots,
\end{equation*}
with $J_2=T$, which is the stress-tensor.  The four-fermion sector contains the scalar and the operator
of spin two at $\Delta=6$, both with nonzero OPE coefficients.  At
$\Delta=8$, it contains one scalar, one primary of spin two, and two primaries
of spin four.  The exact Wick correlator in
\smref{app:majorana-correlator} shows that the scalar, the operator of
spin two, and one linear combination in the spin-four space have nonzero OPE
coefficients.  The independent counting in the same appendix determines the
local multiplicities and identifies the orthogonal spin-four combination
that decouples from this correlator.  It also gives the complete block
decomposition and exact OPE weights below dimension ten.

The same scalar-gap problem was previously studied with the derivative basis up to $\Lambda=55$ \cite{Nakayama:2019jvm}. No kink was observed at
$\Delta_\phi=2$, and an extrapolation gave
$\Delta_{\mathrm{gap}}\gtrsim6.147$. The extracted extremal spectrum also showed no clear integer structure. Our derivative-basis results agree with the previous finite-$\Lambda$ data and extend them to $\Lambda=63$, while the
product-functional bounds converge more rapidly. Our data at larger cutoffs instead lead to a lower extrapolated value.
Independent power-law extrapolations of the finite-basis sequences from the two numerical methods favor a common asymptotic scalar gap near:
\begin{equation}
  \Delta^{\ell=0}_{\mathrm{gap}}\simeq6.086(3),
  \label{eq:majorana-gap}
\end{equation}
approximately $1.4\%$ above the free Majorana value.  This numerical extrapolation disfavors exact
saturation by the free theory.
% although it does not constitute a rigorous exclusion of the limiting value $6$.

\begin{table}[H]
  \centering\small
  \begin{tabular}{l c c c c}
    \toprule
    schematic operator
      & free $(\Delta,\ell)$ & free $\sum_i f_i^2$
      & extremal $(\Delta,\ell)$ & extremal $f^2$ \\
    \midrule
    $T=J_2\sim\psi\gamma\partial\psi$
      & $(3,2)$ & $3/2\;(1.500)$ & $(3.000,2)$ & $1.515$ \\
    $J_4\sim\psi\gamma\partial^3\psi$
      & $(5,4)$ & $35/128\;(0.2734)$ & $(5.082,4)$ & $0.2748$ \\
    $\epsilon'=\epsilon\Box\epsilon$
      & $(6,0)$ & $2\;(2.000)$ & $(6.090,0)$ & $2.047$ \\
    $\epsilon\,\partial\partial\epsilon$
      & $(6,2)$ & $3/4\;(0.7500)$ & $(5.861,2)$ & $0.7109$ \\
    $J_6\sim\psi\gamma\partial^5\psi$
      & $(7,6)$ & $2079/65536\;(0.03172)$
      & $\substack{(7.000,6)\\(7.354,6)}$
      & $\substack{0.01962\\0.01231}$ \\
    $\epsilon\Box^2\epsilon$
      & $(8,0)$ & $9/88\;(0.1023)$ & not resolved & not resolved \\
    $\epsilon\,\partial^2\Box\epsilon$
      & $(8,2)$ & $81/154\;(0.5260)$ & $(7.956,2)$ & $0.5725$ \\
    $\epsilon\,\partial^4\epsilon$
      & $(8,4)$ & $35/192\;(0.1823)$ & $(7.938,4)$ & $0.1727$ \\
    $J_8\sim\psi\gamma\partial^7\psi$
      & $(9,8)$ & $6435/2097152\;(0.003068)$
      & $\substack{(9.022,8)\\(9.603,8)}$
      & $\substack{0.002253\\0.0008500}$ \\
    \bottomrule
  \end{tabular}
  \caption{Comparison of the exact free Majorana OPE with the direct
  $\lambda=12$ boundary solution of the scalar gap problem at
  $\Delta_\phi=2$.  The first column gives schematic representatives, as in
  Fig.~\ref{fig:dphi2-spectrum}, rather than explicit primary projections.
  The free weight column gives the exact total squared OPE coefficient at
  fixed $(\Delta,\ell)$, followed by its decimal approximation in
  parentheses.
  Extremal dimensions are rounded to three decimal places, while extremal
  OPE weights are shown to four significant figures.  The spin-four sector
  at $(8,4)$ has local multiplicity two, but only one linear combination
  couples.  The free current rows of spins six and eight each display two
  nearby extremal levels.  This placement does not identify either level
  with the free current and gives no direct interpretation to the sum of
  their weights.}
  \label{tab:majorana}
\end{table}

%The zeros of the extremal functional give candidate operator dimensions on
%the boundary of the numerical problem.  Their spin and total OPE weight can
%be extracted. 
We also extract the extremal spectrum associated with this bound. For operators degenerate in conformal dimension, a single scalar correlator determines only the sum of their squared OPE coefficients, not the individual coefficients. %At an exact degeneracy, however, a single scalar correlator determines only the total weight and not the basis-dependent individual OPE coefficients.  
We write $f_i=f_{\epsilon\epsilon\mathcal O_i}$. Table~\ref{tab:majorana} reports the summed free weight at fixed
$(\Delta,\ell)$ and lists the extremal spectrum 
 found at finite basis size. Our OPE coefficients are normalized according to the conformal block convention defined in \smref{app:majorana-block-convention}. %Every OPE weight in this comparison multiplies a unit-lightcone block as defined in \smref{app:majorana-block-convention}.  
 The ``free"
entries include every
nonidentity operator below dimension ten that appears in the exact
$\epsilon\times\epsilon$ OPE.  The ``extremal" entries are the direct $\lambda=12$ data, without extrapolation.

% [ht!], not [H]: this figure is nearly a full page, and [H] strands the
% remainder of the preceding page as blank space when it does not fit.
% The ! is required: without it the float exceeds \topfraction and gets
% deferred past the later figures, breaking their numbering order.
\begin{figure}[ht!]
  \centering
  \includegraphics[width=0.90\linewidth]{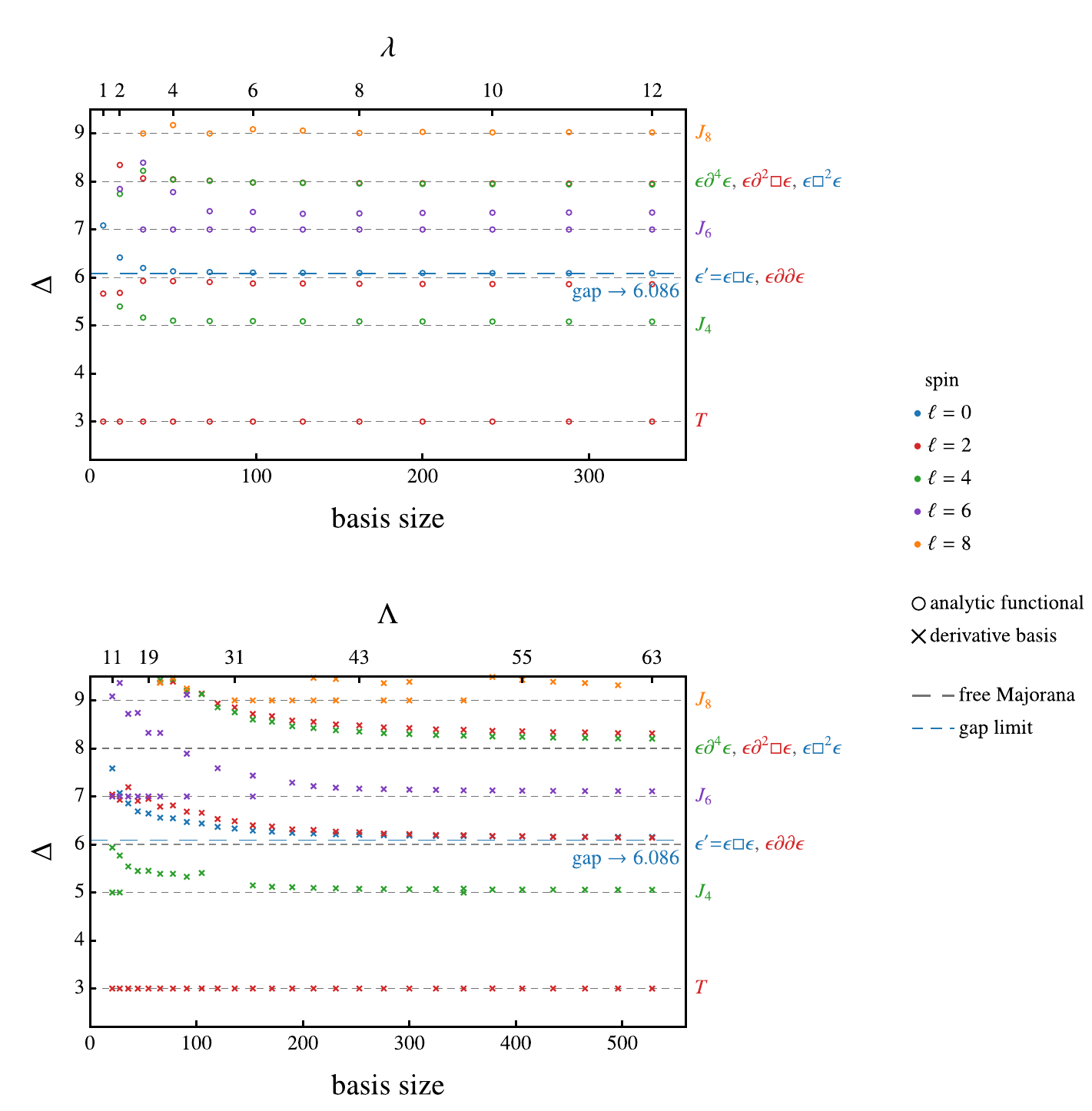}
  \caption{%
    Extremal spectrum of the scalar gap problem at $\Delta_\phi=2$ versus
    basis size, restricted to $\Delta\leq9.5$ and $\ell\leq8$.  Color denotes
    spin.  The upper panel shows the analytic basis through $\lambda=12$ as
    hollow circles, with $\lambda$ on the upper axis.  The lower panel shows
    the derivative basis through $n_{\max}=32$ as crosses, with
    $\Lambda=2n_{\max}-1$ on the upper axis.  Gray dashed lines mark free
    Majorana reference dimensions.  The lines at $\Delta=3,5,7,9$ are the
    conserved current levels $J_2,J_4,J_6,J_8$.  The line at $\Delta=6$
    contains the established four-fermion scalar and operator of spin two.
    The line at $\Delta=8$ contains three four-fermion sectors with nonzero
    total OPE weight.  The spin-four sector has local multiplicity two, of
    which one linear combination couples.  Labels at this dimension identify
    spin sectors rather than explicit primary projections.  The blue dashed
    line marks the estimated scalar gap $6.086$.%
  }
  \label{fig:dphi2-spectrum}
\end{figure}

Several entries lie close to the free reference data, but the complete
pattern is not that of the free theory.  The level at $\Delta=3.000$ has spin
two and can be identified with a stress tensor if the boundary solution is a
local CFT.  Its squared OPE coefficient differs from the free value by about $1\%$.
%one percent.  
The leading scalar has $\Delta=6.090$ and $f^2=2.047$, compared
with the free values $6$ and $2$, while the next level of spin two lies at
$\Delta=5.861$ rather than at the free value $6$.

A more structural mismatch occurs near dimension eight.  The extremal spectrum by analytic functional resolves operators of spins four and two at $7.938$ and $7.956$, but
no stable scalar operator near dimension 8.  %level.  
The exact free correlator assigns the strictly
positive weight $9/88$ to the scalar at $(\Delta,\ell)=(8,0)$, so its absence
cannot be attributed to a counting ambiguity.  The next resolved extremal
scalar lies instead at $\Delta=9.747$.  Although its weight
$f^2=0.09943$ is close to $9/88$, the displacement in dimension prevents an
identification with the free scalar. We also compare the data for higher-spin operators. The spin-4
extremal operator occurs at $\Delta=5.082$, rather than at the conserved current ($J_4$) value $\Delta=5$. In both the
spin-6 and spin-8 sectors, the extremal spectrum contains two operators near the corresponding free-current dimension. The sums of their squared OPE coefficients agree with the free-current OPE coefficients at the $0.6\%$ and $1.1\%$ levels, respectively. Since the operators occur at different dimensions, this agreement does not give an operator-by-operator
match. Moreover, the exact free OPE contains no second spin-6 operator below dimension ten, as shown in \smref{app:majorana-correlator}.

The fact that $J_4$ does not lie at the conserved value, together with the
additional higher-spin operators and the absence of a scalar at $(8,0)$,
shows that the extremal spectrum does not coincide with that of the free
Majorana theory. If this solution corresponds to a local unitary CFT, its
low-energy data would be close to those of an interacting deformation of the
free Majorana theory. A mixed-correlator bootstrap, supplemented by mild
assumptions on the spectrum, may isolate this solution in an island, as in the
Ising case.

%Crossing alone does not establish the existence or identity of such a CFT.

\subsection{Mean field theory and the spin-two gap}
\label{sec:gff}

We now reconsider the gap maximization problem in the spin-2 sector. %sector of spin two.
 Since the spin is fixed, maximizing the operator dimension is equivalent to
maximizing its twist, $\tau_{\mathrm{gap}}^{(2)}=
  \Delta_{\mathrm{gap}}^{(2)}-2.$
The generalized-free-field correlator, which we also refer to as the mean-field-theory solution, contains the leading double-twist operator
$[\phi\phi]_{0,2}$ with
\begin{equation}
  \tau_{0,2}=2\Delta_\phi,
  \qquad
  \Delta_{0,2}=2\Delta_\phi+2.
  \label{eq:gffspintwogap}
\end{equation}
Although this correlator does not in general describe a local CFT with a stress tensor, it is a unitary crossing-symmetric solution corresponding to the boundary correlator of a free scalar field in $AdS_4$.  It is therefore
admissible in the present optimization, where neither an isolated stress
tensor nor its Ward identity is imposed.

Reference~\cite{Caron-Huot:2020adz} conjectured that the mean-field-theory
value in Eq.~\eqref{eq:gffspintwogap} is the maximal twist gap at every fixed
even spin $\ell\geq2$.  For spin two, the proposed extremal functional has
the schematic form
\begin{equation}
  \Phi_2=\beta_{0,2}
  +\bigl[\text{functionals supported at odd spin}\bigr].
  \label{eq:phitwostructure}
\end{equation}
$\beta_{0,2}$ is the functional dual to GFF spin-2 leading double twist operator. $\Phi_2$ has a simple zero on $[\phi\phi]_{0,2}$ and double zeros on the remaining
mean-field-theory double-twist spectrum.  If its action is nonnegative on
every block allowed by unitarity, with the spin-two sector restricted to
$\tau\geq2\Delta_\phi$, the associated sum rule certifies
\begin{equation}
  \tau_{\mathrm{gap}}^{(2)}\leq2\Delta_\phi.
  \label{eq:spintwobound}
\end{equation}
%Reference~
In \cite{Caron-Huot:2020adz} the authors tested the required positivity numerically
and %found it throughout $d=3$ 
found it to hold in $d=3$ for $\tfrac12\leq\Delta_\phi\leq1$.
%$\tfrac12\leq\Delta_\phi\leq1$.  
As $\Delta_\phi$ increases, the critical
twist below which $\Phi_2$ can become negative rises and eventually crosses
the unitarity bound. In $d=3$ %In $d=3$, Ref.~\cite{Caron-Huot:2020adz} 
locates this crossing only within $1<\Delta_\phi<2$.  Beyond this value of $\Delta_{\phi}$, %crossing, %this particular representative
$\Phi_2$ develops a negative region in a sector of even spin greater
than two.  This does not disprove the bound.  The odd-spin term in
Eq.~\eqref{eq:phitwostructure} is ambiguous, and another functional %representative 
may in principle restore positivity.

Our numerical results nevertheless give strong evidence for mean-field-theory saturation beyond the positivity range established by the explicit functional
$\Phi_2$.  For
$\Delta_\phi=1,\tfrac32,2,\tfrac52,3$, the residual, $|
  \Delta_{\mathrm{gap}}^{(2)}
  -(2\Delta_\phi+2)|$
decreases as a power of the basis size in both numerical approaches, as
shown in \mainref{fig:spin2-convergence}. At $\Delta_\phi=1$ this behavior is consistent with what is expected from $\Phi_2$. %the analytic candidate and its positivity tests. }  
For $\Delta_\phi>1$ it should instead be regarded as numerical evidence for the
conjectured bound, possibly realized by a different representative of the
extremal functional.

\begin{figure}[H]
  \centering
  \includegraphics[width=0.84\linewidth]{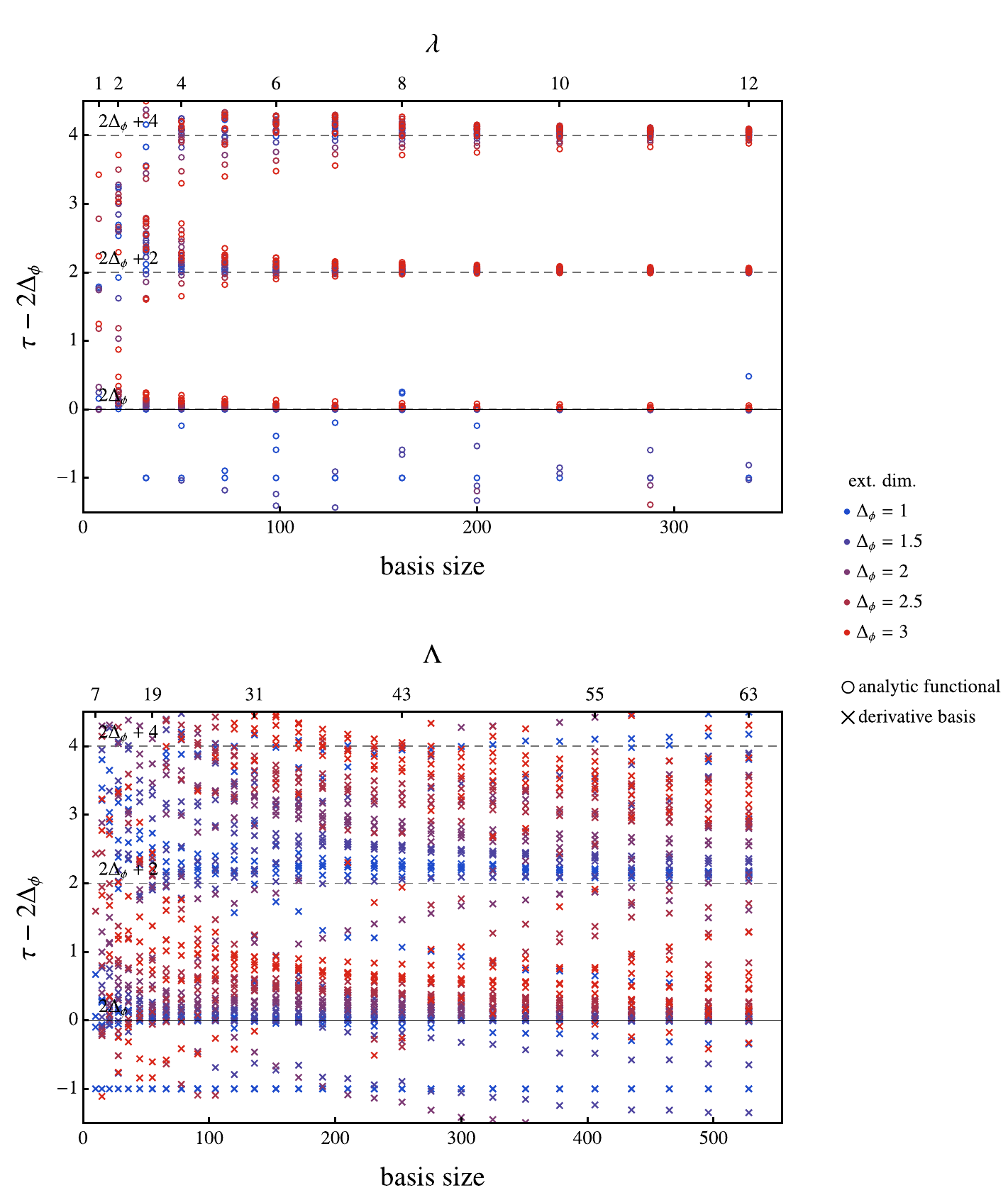}
  \caption{Shifted twists in the extremal solution for
    $\Delta_\phi=1,\tfrac32,2,\tfrac52,3$.  The horizontal lines at
    $\tau-2\Delta_\phi=0,2,4$ mark the first three mean-field-theory double-twist families.  The upper panel shows the analytic functional results as
    hollow circles, with $\lambda$ on the upper axis.  The lower panel shows
    the derivative basis results as crosses, with $\Lambda$ on the upper axis.
    Colors distinguish the external dimensions.  Levels at the unitarity
    boundary in ungapped sectors are numerical artifacts and do not represent
    operators in the candidate mean-field-theory spectrum.}
  \label{fig:twist-collapse}
\end{figure}

The subleading zeros of the numerical extremal functional provide further evidence that the bound is saturated by the GFF solution. % a complementary test.  
Figure \ref{fig:twist-collapse} displays the extracted spectrum in terms of
$\tau-2\Delta_\phi$.  The levels approach the integer bands $0$, $2$, and
$4$, which are the locations of the mean-field-theory towers
$[\phi\phi]_{n,\ell}$.  The analytic functional basis resolves this structure
at substantially smaller basis size than the derivative basis.
This agreement supports the identification of the extremal spectrum with the GFF spectrum. %In view of the ambiguity discussed above, the observed collapse is nontrivial evidence for a mean-field-theory primal solution, but it should not be interpreted as a general uniqueness theorem for the extremal functional.

\begin{figure}[H]
  \centering
  \includegraphics[width=0.7\linewidth]{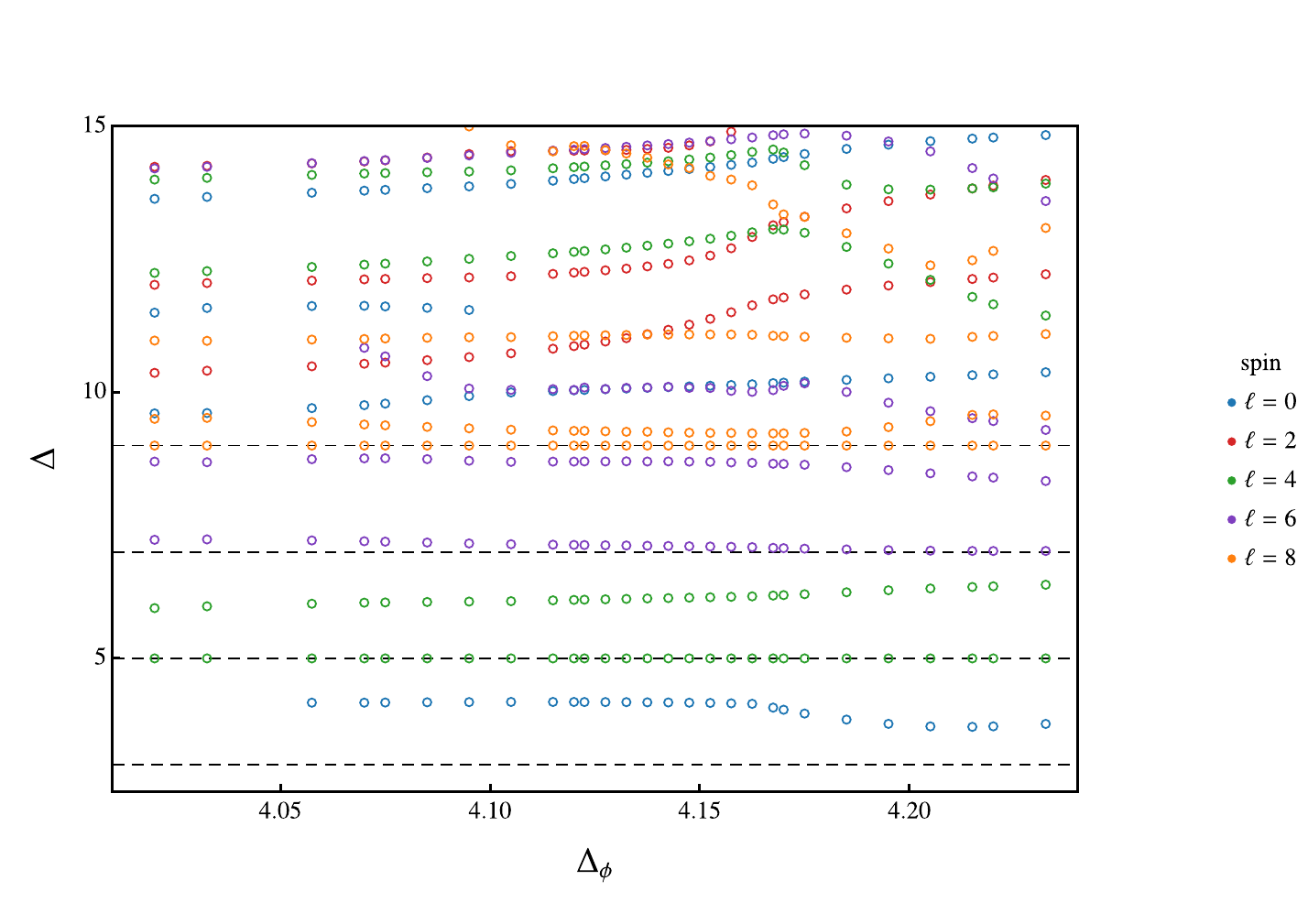}
  \caption{Extremal spectrum across the kink region at $\lambda=12$.
    The reorganization is centered near $\Delta_\phi\simeq4.16$.  Dashed
    lines mark the unitarity bounds for the spinning sectors.  All displayed
    operators have squared OPE coefficient at least $10^{-5}$ times the
    corresponding mean-field-theory value.}
  \label{fig:kink-spectrum}
\end{figure}

 At larger external dimension, a different pattern emerges. In the $\lambda=12$ spectrum shown in Fig.~\ref{fig:spectrum-dphi}, the mean-field-theory double-twist structure begins to break down around $\Delta_\phi\simeq3$. It seems to suggest the extrapolated bound may begin to move smoothly away from the mean-field-theory value for $\Delta_\phi\gtrsim3$.
A sharp kink subsequently develops in \mainref{fig:spin2-bound}. Its location
stabilizes near $\Delta_\phi\simeq4.16$ at the largest analytic cutoffs. To
our knowledge, this is the first observation of this kink. The extremal
spectrum changes sharply at the same location, as shown in
Fig.~\ref{fig:kink-spectrum}. The lowest spin-two operator moves to higher
dimension, several nearby levels come close before separating again, and a
new low-lying scalar appears beyond the kink. This simultaneous change in
several sectors suggests that the solutions on the two sides of the kink are
distinct. We do not know a three-dimensional local CFT corresponding to the
solution beyond the kink. Since we do not impose the existence of a conserved
stress tensor, one possible interpretation is a three-dimensional defect CFT
embedded in a higher-dimensional local CFT, whose defect OPE need not contain
a conserved three-dimensional stress tensor. At present, this interpretation
remains speculative.

\section{Outlook}
 We have constructed and implemented product analytic functionals for the three-dimensional single-correlator bootstrap. At comparable basis size, they give substantially stronger bounds than the derivative basis, with the
difference becoming larger at high external dimension, and reveal features of the boundary that are difficult to resolve with derivative functionals. Of course, the bootstrap program is not limited to determining the data of CFTs already known
from Lagrangian descriptions. We hope that it can also provide evidence for previously unknown CFTs through kinks and other distinguished features of the allowed region. In our earlier work \cite{Ghosh:2023onl}, and again in the present
analysis, such features appear in simple gap-maximization problems. Their appearance should therefore be investigated as a possible indication of new CFTs rather than dismissed as a
numerical artifact. The new kink in the spin-2 bound is one such example. The extremal spectrum
begins to depart from its mean-field-theory organization for
$\Delta_\phi\gtrsim3$ and changes sharply near $\Delta_\phi\simeq4.16$, but we do not presently know a CFT corresponding to the solution at the kink. The same
question applies to the plateaux observed at larger external dimension. Clarifying the nature of these extremal solutions will require higher cutoffs, systematic tracking of operator dimensions and OPE coefficients \cite{Erramilli:2026esf}, and a mixed-correlator bootstrap. Moment observables
may also be useful for following changes in the extremal spectrum and operator decoupling \cite{Chiang:2026moments}. Comparisons with other unexplained kinks in bootstrap bounds may provide further clues \cite{Nakayama:2024magnetization}. This brings us to an immediate target of the product-functional program: extending the present construction to mixed
correlators and theories with global symmetry. The Ising analysis gives a direct motivation for the mixed-correlator
extension. A single correlator reproduces the low-lying Ising data with high accuracy, while its extremal spectrum differs from the Ising spectrum at higher dimensions. Extending the product basis to matrix crossing equations would allow it to be used in precision-island studies and in the bootstrap of
$O(N)$ models \cite{Kos:2013tga,Kos:2014bka,Kos:2016ysd,Chester:2019ifh,He:2020azu,He:2023currents,Reehorst:2024frustrated,Chang:2024whx}.
The same construction can be applied to CFTs with hypercubic,
hypertetrahedral, $MN$, tetragonal, $O(m)\times O(n)$, and
$U(m)\times U(n)$ symmetry
\cite{Stergiou:2018gjj,Kousvos:2018rhl,Kousvos:2019hgc,Stergiou:2019dcv,Henriksson:2020fqi,Kousvos:2021rar,Kousvos:2022ewl,Kousvos:2025ext}. Several of these systems exhibit kinks at comparatively large external dimension, where the faster convergence of the product basis may be especially
useful. Mixed-correlator problems require an optimization framework for matrix-valued positivity. The extension to single-correlator problems with global symmetry is straightforward and will be presented in a separate publication. The dimensional-reduction construction can also be extended beyond three
dimensions. The implementation is most direct in even dimensions, where closed-form expressions for conformal blocks are available. A four-dimensional implementation is currently underway.

The increasing advantage at large external dimension makes conformal gauge
theories a particularly compelling application.  Gauge invariant and
monopole operators often have dimensions well above those of the low lying
Ising operators, precisely where the derivative basis becomes expensive
\cite{Poland:2022qrs,Rychkov:2023wsd,Rychkov:2025boot,He:2023currents,Chester:2025uxb}.
Existing numerical studies have bootstrapped fermionic QED$_3$ using
monopole and adjoint bilinear correlators
\cite{Chester:2016wrc,Chester:2017vdh,Li:2018lyb,Li:2021emd,He:2021sto,Albayrak:2021xtd}.
Scalar QED$_3$ and related deconfined critical points have also been studied
using decoupling conditions, adjoint correlators, and monopole correlators
\cite{He:2021xvg,Manenti:2021elk,Chester:2025uxb}.
Combining analytic functionals with large $N_f$ calculations, perturbative
data, and symmetry constraints could make mixed-correlator studies of
QED$_3$ and deconfined critical points substantially more tractable.  The
long term objective is to isolate their allowed parameter regions and
extract precise CFT data.

Long-range Ising models provide another controlled setting, with perturbative
data, analytic sum rules, and numerical results already available
\cite{Behan:2023lri,Benedetti:2024longrange,Benedetti:2025nzp,Ghosh:2026gku}.
This direction also connects with the extensive bootstrap literature on
boundary and defect correlators
\cite{Padayasi:2021sik,Lanzetta:2022lze,Dey:2024ilw,Cavaglia:2024dkk,Barrat:2024aoa,Lanzetta:2025endpoint}. Since analytic functional bases are known for boundary two-point functions and
real projective space
\cite{Kaviraj:2018tfd,Mazac:2018biw,Giombi:2020xah}, our numerical strategy
could be extended to these settings.
% At large spin, the spectrum approach to generalized free field data is known under broad assumptions \cite{VanRees:2024lightcone}, while the large external dimension regime at fixed spin studied here is complementary. 
The improved convergence of the product basis at large external dimension also makes it well suited to exploring the flat-space $S$-matrix bootstrap through bootstrapping boundary correlators in AdS space \cite{Chang:2023szz,Paulos:2017fhb}.% A distinct limit involving heavy exchanged operators connects dispersive CFT sum rules to flat space scattering \cite{Chang:2023szz} and may provide an interface with direct
%amplitude \red{S-matrix} bootstrap methods \cite{Paulos:2017fhb}.

\section*{Acknowledgements}
We are grateful to Antonio Antunes, Connor Behan, Simon Caron-Huot,
Rajeev Erramilli, Yin-Chen He, Apratim Kaviraj, Ryan Lanzetta,
Dalimil Maz\'{a}\v{c}, Ian Moult, Yu Nakayama, Miguel F.~Paulos,
David Poland, Slava Rychkov, David Simmons-Duffin, Aninda Sinha,
Ning Su, Philine van Vliet, Balt van Rees, Pedro Vieira, Yuan Xin,
Xi Yin, Zahra Zahraee, and Xinan Zhou for helpful discussions and
valuable input. K.G. is supported by the Royal Society under grant
RF\textbackslash{}ERE\textbackslash{}231142.  Z.Z. is supported by Simons
Foundation grant \#994308 for the Simons Collaboration on Confinement and
QCD Strings.

% ---------------- Appendices ----------------
\appendix
\section{Elements of the one-dimensional analytic functionals}
\label{app:oned}
The product functional uplift one-dimensional analytic functionals to higher dimensions. In this
appendix, \(\Delta_\phi\) denotes the one-dimensional external dimension. We
normalize the \(SL(2,\mathbb R)\) block and its two crossing combinations as
\begin{equation}
\begin{aligned}
  G_\Delta(z)
  &=
  z^\Delta\,{}_2F_1(\Delta,\Delta;2\Delta;z),
  \\
  F^{\mp}_\Delta(z)
  &=
  z^{-2\Delta_\phi}G_\Delta(z)
  \mp
  (1-z)^{-2\Delta_\phi}G_\Delta(1-z).
\end{aligned}
\label{eq:oned-vectors}
\end{equation}
Thus \(F^-_\Delta\) is crossing antisymmetric and \(F^+_\Delta\) is crossing
symmetric. Both are analytic on the cut plane
\begin{equation}
  \mathcal{R} \;=\; \mathbb{C}\setminus\bigl((-\infty,0]\cup[1,\infty)\bigr),
  \label{eq:cutplane}
\end{equation}
with branch points at $z=0$ and $z=1$, and crossing is a statement about
them on all of $\mathcal{R}$, not just near \(z=\tfrac12\). In the
conventional numerical bootstrap they are expanded as
\begin{equation}
  F^{\pm}_{\Delta}(z)
  =
  \sum_{n=0}^{\infty}
  \frac{1}{n!}
  \left.
  \partial_z^n F^{\pm}_{\Delta}(z)
  \right|_{z=\frac12}
  \left(z-\frac12\right)^n.
\end{equation}
\placefunctionalfigureearly
They also admit expansions on generalized free spectra
\cite{Mazac:2016qev,Mazac:2018mdx,Mazac:2018ycv,Paulos:2019gtx},
\begin{equation}   \label{eq:master}
  F^{\pm}_{\Delta}(z)
  =
  \sum_n \alpha^{\pm}_n(\Delta) F^{\pm}_{\Delta_n}(z)
  +
  \sum_n \beta^{\pm}_n(\Delta) \partial F^{\pm}_{\Delta_n}(z),
\end{equation}
where
\(\partial F_\Delta^\pm\equiv\partial_\Delta F_\Delta^\pm\), and
\begin{equation}
\begin{aligned}
  \Delta^B_n
  &=
  2\Delta_\phi+2n,
  \\
  \Delta^F_n
  &=
  2\Delta_\phi+2n+1.
\end{aligned}
\end{equation}
Equation~\eqref{eq:master} applies directly to the fermionic basis. For the
bosonic family, it refers to the prefunctionals before the Regge subtraction
described below.

The functionals admit a cut representation, acting through the
discontinuity of the crossing vector along \([1,\infty)\),
\begin{equation}
  \omega\bigl[F^\pm_\Delta(z)\bigr]
  =
  \int_1^\infty\frac{dz}{\pi}\,
  h_\omega(z)\,
  \mathcal I_z F^\pm_\Delta(z),
  \qquad
  \mathcal I_z X(z)
  \equiv
  \lim_{\epsilon\to0^+}
  \frac{X(z+i\epsilon)-X(z-i\epsilon)}{2i},
  \label{eq:oned-cut-appendix}
\end{equation}
which is useful for defining them.  At special external dimensions, the
kernels can be elementary.  For example, at \(\Delta_\phi=\tfrac12\), the
generalized free fermion extremal functional \(\beta^-_{F,0}\), acting on
\(F^-_\Delta\), has the cut kernel
\begin{equation}
  h_{\beta^-_{F,0}}(z)
  \propto
  1-\frac{1}{z(1-z)}
  -\frac{(1-z)(2z^2+z+2)}{2z^2}\log(z-1)
  -\frac{z(2z^2-5z+5)}{2(1-z)^2}\log z,
  \qquad z>1.
\label{eq:beta-minus-half-kernel-example}
\end{equation}
The proportionality reflects an arbitrary positive normalization
\cite{Mazac:2016qev}.  For general external dimension, and for the other
crossing parity, exact kernels are naturally expressed in terms of generalized
hypergeometric functions \cite{Mazac:2018mdx,Paulos:2019gtx}.  They obey the
gluing and falloff conditions required for finiteness and swapping.  The actions
\(\alpha^{\pm}_{n}(\Delta)\) and \(\beta^{\pm}_{n}(\Delta)\) form a dual
basis. Suppressing the sign and family labels, their duality relations are
\begin{equation}
\begin{aligned}
  \alpha_{m}\bigl(\Delta_n\bigr)&=\delta_{mn},
  &
  \partial\alpha_{m}\bigl( \Delta_n\bigr)&=-\delta_{0n}\, c_m,
  \\
  \beta_{m}\bigl(\Delta_n\bigr)&=0,
  &
  \partial\beta_{m}\bigl( \Delta_n\bigr)&=\delta_{mn}-\delta_{0n}\, d_m.
\end{aligned}
  \label{eq:duality}
\end{equation}
For the fermionic basis, \(c_m=d_m=0\). For the bosonic basis acting on
\(F^-_\Delta\), the admissible functionals are obtained from prefunctionals
by
\begin{equation}
\begin{aligned}
  \alpha^-_{B,m}
  &=
  \bar\alpha^-_{B,m}
  -
  c_m\bar\beta^-_{B,0},
  \\
  \beta^-_{B,m}
  &=
  \bar\beta^-_{B,m}
  -
  d_m\bar\beta^-_{B,0},
  \\
  d_m
  &=
  \frac{[(\Delta_\phi)_m]^4}{(m!)^2[(2\Delta_\phi)_m]^2}
  \frac{(4\Delta_\phi-1)_{2m}}
       {(4\Delta_\phi+2m-1)_{2m}},
  \\
  c_m&=\frac{1}{2}\partial_m d_m.
\end{aligned}
\label{eq:bosonic-subtraction}
\end{equation}
Here \((a)_m\) is the rising Pochhammer symbol, and \(\partial_m d_m\) is
evaluated after analytic continuation away from integer \(m\). Since
\(d_0=1\), the subtracted \(\beta^-_{B,0}\) vanishes identically. The
admissible bosonic basis contains \(\alpha^-_{B,m}\) for \(m\geq0\) and
\(\beta^-_{B,m}\) for \(m\geq1\). In this paper we use this bosonic basis on
\(F^-_\Delta\) and the fermionic basis on \(F^+_\Delta\).

Direct numerical integration of
Eq.~\eqref{eq:oned-cut-appendix} is too slow for producing the
functional tables used in this work.  Instead, we use the relation between
functional actions on crossing antisymmetric block and coefficient of the conformal block decomposition of crossing
symmetric sums of exchange Witten diagrams in \(AdS_2\), supplemented by the
contact diagrams required by the Regge subtraction
\cite{Mazac:2018ycv,Ferrero:2019luz}. Similarly, the relation between functional actions on crossing symmetric block and the coefficient of conformal block decomposition of the appropriate combination of exchange Witten diagram and contact term was derived in \cite{Kaviraj:2021cvq}. Following the Witten diagram recursion
relations of Ref.~\cite{Zhou:2018sfz}, we apply the \(AdS_2\) Casimir equation
to reduce an exchange diagram to contact diagrams with known conformal block
coefficients. Matching coefficients then gives the recursion relations for
the functional actions used in our numerical implementation
\cite{Ghosh:2021ruh}.

The payoff of this construction is its domain of convergence, illustrated in
Fig.~\ref{fig:comparison} at equal basis size \(\Lambda=10\), with ten Taylor
powers and ten double-twist elements. The Taylor truncation converges only
for \(|z-\tfrac12|<\tfrac12\), whereas the analytic reconstruction remains
accurate throughout the cut plane \(\mathcal R\). Since an optimized
functional probes crossing beyond the Taylor disk
\cite{Mazac:2016qev,CastedoEcheverri:2016fxt}, the
analytic basis is naturally adapted to this global structure. 

  \begin{functionalfigure}
  \centering
  \includegraphics[width=\functionalfigwidth]{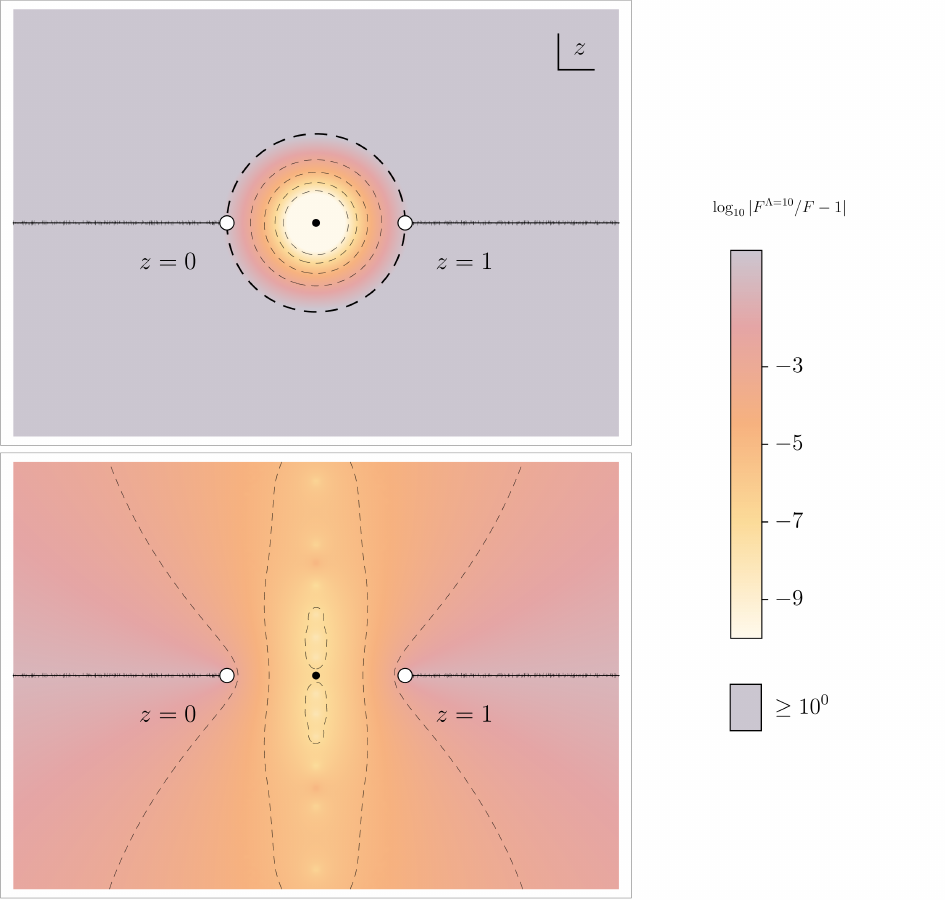}
  \caption{%
    Truncation error
    \(\log_{10}\bigl|F_{\mathrm{trunc}}(z)/F(z)-1\bigr|\) for two
    reconstructions at equal basis size \(\Lambda=10\): ten Taylor powers
    in the top panel and ten double-twist elements in the bottom panel.
    The Taylor series about $z=\tfrac12$ converges only inside
    $|z-\tfrac12|<\tfrac12$, bounded by the branch points $z=0$ and $z=1$;
    gray denotes error $\geq1$. The analytic reconstruction is accurate
    throughout the cut plane $\mathcal{R}$ of
    Eq.~\eqref{eq:cutplane}. Thin dashed curves are contours of constant
    error.%
  }
  \label{fig:comparison}
\end{functionalfigure}

A natural alternative is to expand in the Zhukovsky variable $
z(y)=\frac{(1+y)^2}{2(1+y^2)},$
which maps the cut plane $\mathcal R$ to the unit disk, with $z=\tfrac12$ corresponding to $y=0$. A Taylor series in y therefore converges throughout $\mathcal R$, unlike the Taylor series in $(z-\tfrac12)$. This improvement of the convergence domain should, however, be distinguished from the choice of finite-dimensional approximation space. At fixed $\Lambda$, the Zhukovsky expansion spans the generic polynomial space ({1,y,$\ldots$,$y^{\Lambda-1}$}), whereas the analytic-functional reconstruction spans functions adapted to the double-twist spectrum and already carrying the appropriate branch-point behavior. The two descriptions may become complete as $\Lambda\to\infty$, but their finite-dimensional subspaces are different, explaining why the analytic basis can give a much more accurate reconstruction at the same basis size. In the dual formulation, finite sets of z- and y-derivatives span the same functional space, while an analytic functional corresponds to a nontrivial infinite resummation of infinite y-derivatives.

\section{Expansion of the full crossing vector in the product basis}
\label{app:product}

The one dimensional functional decompositions in
\maineqref{eq:master} %induce 
lead to an expansion of the full crossing vector in
the product basis in any spacetime dimension \(d\geq2\):
\begin{small}
\begin{equation}
\begin{split}
  F^{d}_{\Delta,\ell}(z,\bar z\mid\Delta_\phi)
  =
  \sum_{m,n}\bigg\{
  &\left(\alpha^-_m\otimes\alpha^+_n\right)
  \left[F^d_{\Delta,\ell}(z,\bar z\mid\Delta_\phi)\right]
  \frac12\left[
    F^-_{\Delta_m}(z)F^+_{\Delta_n}(\bar z)
    +
    F^-_{\Delta_m}(\bar z)F^+_{\Delta_n}(z)
  \right]
  \\
  &+
  \left(\beta^-_m\otimes\alpha^+_n\right)
  \left[F^d_{\Delta,\ell}(z,\bar z\mid\Delta_\phi)\right]
  \frac12\left[
    \partial F^-_{\Delta_m}(z)F^+_{\Delta_n}(\bar z)
    +
    \partial F^-_{\Delta_m}(\bar z)F^+_{\Delta_n}(z)
  \right]
  \\
  &+
  \left(\alpha^-_m\otimes\beta^+_n\right)
  \left[F^d_{\Delta,\ell}(z,\bar z\mid\Delta_\phi)\right]
  \frac12\left[
    F^-_{\Delta_m}(z)\partial F^+_{\Delta_n}(\bar z)
    +
    F^-_{\Delta_m}(\bar z)\partial F^+_{\Delta_n}(z)
  \right]
  \\
  &+
  \left(\beta^-_m\otimes\beta^+_n\right)
  \left[F^d_{\Delta,\ell}(z,\bar z\mid\Delta_\phi)\right]
  \frac12\left[
    \partial F^-_{\Delta_m}(z)\partial F^+_{\Delta_n}(\bar z)
    +
    \partial F^-_{\Delta_m}(\bar z)\partial F^+_{\Delta_n}(z)
  \right]
  \bigg\}.
  \label{eq:full-crossing-reconstruction}
\end{split}
\end{equation}
\end{small}
All one dimensional vectors and functionals in this equation have external
dimension \(\Delta_\phi/2\).  The derivatives act with respect to the exchanged
dimension and are evaluated at \(\Delta_m\) or \(\Delta_n\).  The product functional
actions are defined in \maineqref{eq:product-cut-definition}.

As a sanity check, in two dimensions the crossing vector factorizes into product of
one dimensional crossing symmetric and antisymmetric vectors,
\begin{small}
\begin{equation}
  F^{d=2}_{\Delta,\ell}(z,\bar z\mid\Delta_\phi)
  =
  \frac14\bigl[
  F^-_{\delta_-}(z)F^+_{\delta_+}(\bar z)
  +
  F^-_{\delta_+}(z)F^+_{\delta_-}(\bar z)
  +
  F^-_{\delta_-}(\bar z)F^+_{\delta_+}(z)
  +
  F^-_{\delta_+}(\bar z)F^+_{\delta_-}(z)
  \bigr],
  \label{eq:full-2d-crossing-factorization}
\end{equation}
\end{small}
where \(\delta_\pm=(\Delta\pm\ell)/2\) and every one dimensional vector has
external dimension \(\Delta_\phi/2\).  Inserting this factorized form into
the product action in \maineqref{eq:product-cut-definition} and using the
reflection properties of \(F^\mp\) under \(\bar z\mapsto1-\bar z\)
immediately reproduces the factorized expression in
\maineqref{eq:product-factorization}, and the expansion in
Eq.~\eqref{eq:full-crossing-reconstruction} follows term by term from the
one dimensional decompositions in \maineqref{eq:master}.

For \(d=3\), the coefficients in
Eq.~\eqref{eq:full-crossing-reconstruction} can be evaluated through
dimensional reduction,
\begin{equation}
  F^{d=3}_{\Delta,\ell}(z,\bar z\mid\Delta_\phi)
  =
  \sum_{q,s}
  A_{q,s}(\Delta,\ell)
  F^{d=2}_{\Delta+2q,s}(z,\bar z\mid\Delta_\phi).
  \label{eq:full-reconstruction-dimensional-reduction}
\end{equation}
Applying the product functionals to this relation reproduces
\maineqref{eq:3d-product-action}.

A different analytic functional basis was proposed in
\cite{Mazac:2019shk}, dual to the double twist spectrum in general
dimension:
\begin{equation}
  F^{d=3}_{\Delta,\ell}(z,\bar z\mid\Delta_\phi)
  =
  \sum_{n,J}
  \alpha_{n,J}(\Delta,\ell)\,
  F^{d=3}_{\Delta_{n,J},J}(z,\bar z\mid\Delta_\phi)
  +
  \sum_{n,J}
  \beta_{n,J}(\Delta,\ell)\,
  \partial F^{d=3}_{\Delta_{n,J},J}(z,\bar z\mid\Delta_\phi).
  \label{eq:gff-dual-expansion}
\end{equation}
The corresponding functionals diagonalize the generalized-free-field
spectrum.  However, they do not possess the positivity properties required
for a direct numerical implementation.  Expanding each
\(F^{d=3}_{\Delta_{n,J},J}\) and its derivative in the product basis of
Eq.~\eqref{eq:full-crossing-reconstruction} relates the two functional
constructions.  In particular, each product functional is represented as an
infinite linear combination of the functionals constructed in \cite{Mazac:2019shk}.
This infinite recombination allows the product-functional basis to exhibit
substantially better asymptotic positivity properties.

\section{Conformal block normalization}
\label{app:majorana-block-convention}

This appendix records the conformal block normalization used in this paper.  We use the unit-lightcone
normalization of \cite{Simmons-Duffin:2016wlq}.  As $z\to0$ at fixed
$\bar z$,
\begin{align}
  G_{\Delta,\ell}(z,\bar z)
  &=z^{(\Delta-\ell)/2}k_{\Delta+\ell}(\bar z)
    +O\!\left(z^{(\Delta-\ell)/2+1}\right),
  \label{eq:majorana-unit-lightcone-block}\\
  k_\beta(x)
  &=x^{\beta/2}
    {}_2F_1\!\left(\frac\beta2,\frac\beta2;\beta;x\right).
\end{align}
Thus the coefficient of
$z^{(\Delta-\ell)/2}\bar z^{(\Delta+\ell)/2}$ in the leading monomial is
one.  This is also the convention used for $f_{\mathrm{phys}}^2$ in
Eq.~\eqref{eq:analytic-ope-normalization}.

\section{Improving the asymptotics}
\label{app:asympt}
As discussed in the main text, dimensional reduction of the $3d$ conformal
block, combined with the factorization of the $2d$ crossing vector into $+$
and $-$ type $1d$ crossing vectors, expresses the $3d$ functional action in
\maineqref{eq:3d-product-action} as
a %level 
sum over the $2d$ multiplets, %tower,
\begin{equation}
(\omega^-_m\otimes\omega^+_n)
\!\left[
F^{d=3}_{\Delta,\ell}
\right]
=
\sum_{q=0}^{\infty}
\sum_{s=\ell,\ell-2,\ldots,\ell\,{\rm mod}\,2}
A_{q,s}(\Delta,\ell)\,
\frac{1}{2}
\left(
\omega_m^-(\delta^-_{q,s}\mid\delta_\phi)\,
\omega_n^+(\delta^+_{q,s}\mid\delta_\phi)
+
(s\to-s)
\right),
\label{eq:asympt-recap}
\end{equation}
where $\delta^\pm_{q,s}=(\Delta+2q\pm s)/2$ and $\delta_\phi=\Delta_\phi/2$
are the one-dimensional arguments associated with the $2d$ crossing vector  $F^{d=2}_{\Delta+2q,s}$. Throughout this appendix we
abbreviate the summand as
$\omega^-_m\otimes\omega^+_n(\Delta+2q,s)\equiv
(\omega^-_m\otimes\omega^+_n)\bigl[F^{d=2}_{\Delta+2q,s}\bigr]$.

The direct evaluation of the infinite $q$-sum to desired precision can be slow at large $\Delta$, because the
summand falls off only polynomially at large $q$ and the numerator also grows with $\Delta$. To accelerate the
evaluation, we determine the large-$q$ expansion of the complete summand,
evaluate the exact expression up to a moderate cutoff $q_{\max}$, and
replace the remaining infinite tail by its asymptotic expansion. Since the
terms in this expansion are inverse powers of $\Delta+2q$, the tail can be
summed analytically in terms of the Hurwitz zeta function.

For each fixed value of $s$, we therefore expand
\begin{equation}
A_{q,s}(\Delta,\ell)\,
\omega^-_m\otimes \omega^+_n(\Delta+2q,s)
\end{equation}
at large $q$, keeping $\Delta,\ell,\Delta_\phi$ and the functional labels
fixed. It is convenient to organize this expansion in powers of $\Delta+2q$ rather than directly in powers of $q$. The resulting asymptotic expansion
takes the form
\begin{equation}
A_{q,s}(\Delta,\ell)\,
\omega^-_m\otimes \omega^+_n(\Delta+2q,s)
\sim
\sum_{r=0}^{\infty}
\frac{c_r}{(\Delta+2q)^{4\Delta_\phi+6+r}}.
\label{eq:complete-asymptotic-expansion}
\end{equation}
The coefficients $c_r$ depend on $\Delta,\ell,\Delta_\phi$, on $s$, and on
the chosen product functional, but are independent of $q$. To obtain these coefficients, we separately expand the dimensional-reduction
coefficient $A_{q,s}(\Delta,\ell)$ and the two one-dimensional functional
actions entering the product. The building blocks of the latter take the
schematic form
\begin{equation}
\omega_m^{\pm}[F^{\pm}_{\Delta}(z)]
\sim
N_{\Delta}
\left(
R_{\Delta}
+
\frac{\Gamma[\cdots]}{\Gamma[\cdots]}
L(a,b,c,d;e;f,g)
\right),
\label{eq:functional-schematic}
\end{equation}
where $R_{\Delta}$ is a rational function of $\Delta$ and
$\Gamma[\cdots]$ denotes a product of Gamma functions whose arguments
depend on $\Delta$. The function $L(a,b,c,d;e;f,g)$ is defined on the Saalsch\"utzian
hyperplane
\begin{equation}
e+f+g-a-b-c-d=1.
\end{equation}
It can be expressed in terms of a very-well-poised ${}_7F_6(1)$ function.
For a ${}_7F_6$ with upper parameters $a_1,\ldots,a_7$ and lower parameters
$b_1,\ldots,b_6$, very-well-poisedness means that
\begin{equation}
1+a_1
=
a_2+b_1
=
a_3+b_2
=
\cdots
=
a_7+b_6,
\end{equation}
together with
\begin{equation}
a_2=1+\frac{a_1}{2},
\qquad
b_1=\frac{a_1}{2}.
\end{equation}
In the present case,
\begin{equation}
\begin{split}
L(a,b,c,d;e;f,g)
={}&
\frac{\Gamma(1+d+g-e)}
{\pi\,
\Gamma(g)\Gamma(1+g-e)\Gamma(f-d)
\Gamma(1+a+d-e)\Gamma(1+b+d-e)\Gamma(1+c+d-e)}
\\
&\times
{}_7F_6\left[
\begin{matrix}
d+g-e,\;
1+\dfrac{d+g-e}{2},\;
g-a,\;
g-b,\;
g-c,\;
d,\;
1+d-e
\\[2mm]
\dfrac{d+g-e}{2},\;
1+a+d-e,\;
1+b+d-e,\;
1+c+d-e,\;
1+g-e,\;
g
\end{matrix}
;1
\right].
\end{split}
\label{eq:L-7F6}
\end{equation}
The parameters appearing in our functional actions are
\begin{equation}
\begin{aligned}
a&=\delta_\phi+\frac{\Delta+\ell}{2}-h,
\quad
b=c
=\delta_\phi+\frac{\Delta+\ell}{2}-h-s+1,\quad
d=1-\delta_\phi+\frac{\Delta-\ell}{2}+m,
\\
e&=\delta_\phi+\frac{\Delta+\ell}{2}-h-2s+2,\quad
f=\delta_\phi+\frac{\Delta+\ell}{2}-h+1,\quad
g=\Delta-h+m+1.
\end{aligned}
\label{eq:L-parameters}
\end{equation}
For example, for the $\beta_{B}^{-}$ functionals, these expressions are evaluated at
\begin{equation}
h=\frac{1}{2},
\qquad
\ell=0,
\qquad
s=\delta_\phi.
\label{eq:betaB-parameters}
\end{equation}

 For general $\Delta_\phi$, the large-$\Delta$ expansion of
Eq.~\eqref{eq:L-7F6} is not manifest in its original representation.
However, the $L$-function is invariant under the action of the Coxeter group
$W(D_5)$ \cite{mishev2012coxeter}. We use this invariance to choose an
equivalent representation in which the hypergeometric function can be
expanded directly in ratios of Pochhammer symbols. In this representation,
successive terms in the hypergeometric series are suppressed by additional
powers of $\Delta^{-2}$.

We first expand the rational and Gamma function prefactors in
Eq.~\eqref{eq:functional-schematic}. We then expand the transformed
${}_7F_6(1)$ term by term and combine the resulting expressions to obtain
the large-dimension expansion of each one-dimensional functional action.
The corresponding arguments $\delta^\pm_{q,s}$ are then substituted and the
result is re-expanded at large q. Finally, we multiply the two
one-dimensional expansions, symmetrize under
$s\to-s$, and combine the result with the large-$q$
expansion of $A_{q,s}(\Delta,\ell)$. This determines the coefficients
$c_r$ in Eq.~\eqref{eq:complete-asymptotic-expansion}. The numerical procedure is then straightforward. We evaluate the exact
summand up to $q=q_{\max}$. For $q>q_{\max}$, we replace the exact summand
by its asymptotic expansion truncated at order $r_{\max}$. Thus,
\begin{equation}
\begin{split}
&
\sum_{q=0}^{\infty}
A_{q,s}(\Delta,\ell)\,
\omega^-_m\otimes \omega^+_n(\Delta+2q,s)
\\
&\qquad\simeq
\sum_{q=0}^{q_{\max}}
A_{q,s}(\Delta,\ell)\,
\omega^-_m\otimes \omega^+_n(\Delta+2q,s)
+
\sum_{r=0}^{r_{\max}}
\sum_{q=q_{\max}+1}^{\infty}
\frac{c_r}{(\Delta+2q)^{4\Delta_\phi+6+r}}.
\end{split}
\label{eq:asymptotic-tail}
\end{equation}
The low-lying part of the dimensional-reduction sum is therefore evaluated
using the exact functional actions, while only the large-$q$ tail is
replaced by its asymptotic approximation.

Each term in the tail can be summed analytically using
\begin{equation}
\sum_{q=q_{\max}+1}^{\infty}
\frac{1}{(\Delta+2q)^p}
=
2^{-p}\,
\zeta\left(
p,\frac{\Delta}{2}+q_{\max}+1
\right),
\label{eq:hurwitz-tail}
\end{equation}
where $\zeta(x,y)$ is the Hurwitz zeta function. Therefore,
\begin{equation}
\begin{split}
&
\sum_{q=q_{\max}+1}^{\infty}
\sum_{r=0}^{r_{\max}}
\frac{c_r}{(\Delta+2q)^{4\Delta_\phi+6+r}}
\\
&\qquad=
\sum_{r=0}^{r_{\max}}
2^{-(4\Delta_\phi+6+r)}c_r\,
\zeta\left(
4\Delta_\phi+6+r,
\frac{\Delta}{2}+q_{\max}+1
\right).
\end{split}
\label{eq:hurwitz-asymptotic-tail}
\end{equation}

The two parameters controlling the approximation are $q_{\max}$, which
determines how many terms are evaluated exactly, and $r_{\max}$, which
determines the number of terms retained in the asymptotic expansion. We
increase both until the resulting functional action is stable at the
required numerical precision.This gives us a practical way to evaluate the three-dimensional functional action to the desired precision, which would otherwise be a very difficult numerical task.

\section{The free Majorana OPE and local primary spectrum}
\label{app:majorana-correlator}

This appendix gives two complementary descriptions of the free Majorana
$\epsilon\times\epsilon$ OPE.  In the first subsection we evaluate the exact
%Wick 
correlator using Wick contraction and extract the OPE data used in the comparison with the
numerical extremal solution.  The conformal block normalization is stated
separately in \smref{app:majorana-block-convention}. % When several primaries share the same $(\Delta,\ell)$, the correlator determines only the total weight
When there is a degeneracy in the spectrum the correlator determines only the total OPE coefficient
$\sum_i f_{\epsilon\epsilon\mathcal O_i}^{\,2}$, not the multiplicity or the
basis-dependent individual coefficients.  In the second subsection we use
an independent character computation to remove descendants and null states
and to list the complete bosonic local primary spectrum through $\Delta=10$.
Together, the two descriptions determine which local sectors occur in this
OPE and make the degeneracy at $(\Delta,\ell)=(8,4)$ explicit.

\subsection{Exact correlator and OPE data}

Let $\psi_\alpha$ be a two component Majorana fermion in three dimensions,
with
\begin{equation}
  \Delta_\psi=1,
  \qquad
  \gamma^\mu\partial_\mu\psi=0,
  \qquad
  \epsilon=\psi^\alpha\psi_\alpha,
  \qquad
  \Delta_\epsilon=2.
\end{equation}
We normalize $\epsilon$ by
\begin{equation}
  \langle\epsilon(x)\epsilon(0)\rangle=|x|^{-4}
\end{equation}
and define
\begin{equation}
  \langle
    \epsilon(x_1)\epsilon(x_2)\epsilon(x_3)\epsilon(x_4)
  \rangle
  =\frac{\mathcal G(u,v)}{x_{12}^4x_{34}^4},
  \qquad
  u=z\bar z,
  \qquad
  v=(1-z)(1-\bar z).
\end{equation}

\paragraph{Exact sector decomposition}

In the $\epsilon\times\epsilon$ OPE, the Wick contractions are naturally
organized by the number of uncontracted fermion fields.  Contracting two,
one, or no fermion pairs between the two external bilinears produces
contributions with, respectively, zero, two, or four uncontracted fermion
fields, corresponding to the identity, two-fermion, and four-fermion
sectors.  Accordingly,
\begin{equation}
  \mathcal G=1+\mathcal G_{\psi^2}+\mathcal G_{\psi^4}.
  \label{eq:majorana-exact-sector-sum}
\end{equation}
A direct Pfaffian evaluation of the Wick contractions gives
\begin{align}
  \mathcal G_{\psi^2}(z,\bar z)
  &=\frac{\sqrt u}{v^{3/2}}
    \left[z+\bar z-2u-v^{3/2}(z+\bar z)\right],
  \label{eq:majorana-exact-bilinear}\\
  \mathcal G_{\psi^4}(z,\bar z)
  &=\frac{u^2}{v^3}
    \left[v^{3/2}-(1-z)\right]
    \left[v^{3/2}-(1-\bar z)\right].
  \label{eq:majorana-exact-quartic}
\end{align}
The identity in Eq.~\eqref{eq:majorana-exact-sector-sum} has squared OPE
coefficient one.

The leading lightcone term of the two-fermion sector is
\begin{equation}
  \mathcal G_{\psi^2}
  =z^{1/2}\bar z^{3/2}
    \left[(1-\bar z)^{-3/2}-1\right]+O(z^{3/2}).
  \label{eq:majorana-bilinear-lightcone}
\end{equation}
Since every conserved current has twist one, comparison with
Eq.~\eqref{eq:majorana-unit-lightcone-block} gives
\begin{equation}
  \bar z^{3/2}\left[(1-\bar z)^{-3/2}-1\right]
  =\sum_{\substack{\ell\geq2\\ \ell\ \mathrm{even}}}
    a_\ell^{(2)}k_{2\ell+1}(\bar z).
  \label{eq:majorana-current-generating-identity}
\end{equation}
Expanding both sides about $\bar z=0$ and subtracting the lower spin blocks
successively yields
\begin{equation}
  a_2^{(2)}=\frac32,
  \qquad
  a_4^{(2)}=\frac{35}{128},
  \qquad
  a_6^{(2)}=\frac{2079}{65536},
  \qquad
  a_8^{(2)}=\frac{6435}{2097152}.
  \label{eq:majorana-current-weights}
\end{equation}
These are the exact squared OPE coefficients of $T,J_4,J_6,J_8$.

Matching the OPE expansion of Eq.~\eqref{eq:majorana-exact-quartic} directly
in the convention of Eq.~\eqref{eq:majorana-unit-lightcone-block} gives
\begin{align}
  \mathcal G_{\psi^4}
  ={}&2g_{6,0}+\frac34g_{6,2}
  +\frac{9}{88}g_{8,0}
  +\frac{81}{154}g_{8,2}
  +\frac{35}{192}g_{8,4}
  \nonumber\\
  &+\text{contributions from primaries with $\Delta\geq10$}.
  \label{eq:majorana-quartic-unit-blocks}
\end{align}

\paragraph{Complete OPE data below dimension ten}

Combining the two-fermion and four-fermion contributions gives
Table~\ref{tab:majorana-exact-ope}.
Every displayed number is the exact coefficient multiplying the unit-lightcone block, and hence is directly comparable with the FunBoot output in
the same convention.
\begin{table}[H]
  \centering
  \begin{tabular}{l c c c}
    \toprule
    sector & representative & $(\Delta,\ell)$
      & $\displaystyle\sum_i f_{\epsilon\epsilon\mathcal O_i}^{\,2}$ \\
    \midrule
    $\psi^2$ & $T=J_2$ & $(3,2)$ & $3/2$ \\
    $\psi^2$ & $J_4$ & $(5,4)$ & $35/128$ \\
    $\psi^4$ & $\epsilon\Box\epsilon$ & $(6,0)$ & $2$ \\
    $\psi^4$ & $\epsilon\,\partial\partial\epsilon$ & $(6,2)$ & $3/4$ \\
    $\psi^2$ & $J_6$ & $(7,6)$ & $2079/65536$ \\
    $\psi^4$ & $\epsilon\Box^2\epsilon$ & $(8,0)$ & $9/88$ \\
    $\psi^4$ & $\epsilon\,\partial^2\Box\epsilon$ & $(8,2)$ & $81/154$ \\
    $\psi^4$ & $\epsilon\,\partial^4\epsilon$ & $(8,4)$ & $35/192$ \\
    $\psi^2$ & $J_8$ & $(9,8)$ & $6435/2097152$ \\
    \bottomrule
  \end{tabular}
  \caption{Exact nonidentity OPE data in the free Majorana
  $\epsilon\times\epsilon$ correlator for $\Delta<10$.  The final column is
  the sum over all primaries with the displayed dimension and spin.}
  \label{tab:majorana-exact-ope}
\end{table}

In particular, the coefficient $35/192$ at $(\Delta,\ell)=(8,4)$ is the sum of OPE coefficients of all the degenrate operators.
%weight.
The correlator alone does not determine how many primaries share
these quantum numbers or how the weight is distributed among them.  The
independent count in the next subsection resolves this multiplicity.
Equation~\eqref{eq:majorana-exact-sector-sum} also shows that operators
containing six or more fermion fields cannot occur in this OPE.

\subsection{Local primary spectrum through dimension ten}
\label{app:majorana-primary-counting}

We now count local primaries independently of the correlator.  Since
$\epsilon\times\epsilon$ is even under fermion parity, only operators with an
even number of fermion fields are relevant.  The result below is complete in
this bosonic sector through $\Delta=10$.  It also includes operators whose OPE
coefficient in the particular $\epsilon\times\epsilon$ correlator vanishes.

\paragraph{One fermion states and the Pauli gap}

By the state-operator correspondence \cite{Simmons-Duffin:2016gjk}, a field
inserted at the origin creates a state in radial quantization.  Let
$\psi_\alpha$ be a two component Majorana spinor with $\Delta_\psi=1$.  At
derivative level $n$, the independent components of
\begin{equation}
  \partial_{(\mu_1}\cdots\partial_{\mu_n)}\psi_\alpha
\end{equation}
are symmetric and traceless in their vector indices and obey the gamma trace
condition.  The ordinary trace vanishes by $\partial^2\psi=0$, while the gamma
trace vanishes by the Dirac equation.  The surviving representation has
\begin{equation}
  j_n=n+\frac12,
  \qquad
  \Delta_n=n+1,
  \qquad
  \dim(j_n)=2n+2.
  \label{eq:majorana-modes}
\end{equation}
Thus level zero contains two states of spin $1/2$, while level one contains
four states of spin $3/2$.  Here a one fermion state is an independent
derivative component in radial quantization, not a momentum mode.

Local products form exterior products of these states.  Fermi statistics
allows each state to occur only once.  A product of four fermions must use
both level zero states and at least two level one states, so
\begin{equation}
  \Delta_{\psi^4}\geq1+1+2+2=6.
  \label{eq:majorana-pauli-gap}
\end{equation}
This filling argument gives the Pauli gap $\Delta_{\epsilon'}=6$.  The two
states at this dimension form a scalar and an operator of spin two.

\paragraph{Character construction}

Introduce fugacities $q$, $z$, and $a$ for scaling dimension, the Cartan
generator $J_3$ of $\operatorname{Spin}(3)$, and fermion number.  We use
\begin{equation}
  \chi_j(z)
  =\operatorname{Tr}_{j}z^{J_3}
  =\sum_{m=-j}^{j}z^m
  =\frac{z^{j+\frac12}-z^{-j-\frac12}}
  {z^{\frac12}-z^{-\frac12}},
  \qquad
  \chi_j(1)=2j+1.
\end{equation}
Equation~\eqref{eq:majorana-modes} gives the one fermion character
\begin{align}
  f(q,z)
  &=\sum_{n=0}^{\infty}q^{n+1}\chi_{n+\frac12}(z)
  \notag\\
  &=\frac{1}{z^{\frac12}-z^{-\frac12}}
  \left(\frac{qz}{1-qz}-\frac{q/z}{1-q/z}\right).
  \label{eq:majorana-single-character}
\end{align}
The fermionic Fock space character is the plethystic exponential
\cite{Benvenuti:2006qr},
\begin{align}
  Z(a,q,z)
  &=\prod_{n=0}^{\infty}
    \prod_{m=-n-\frac12}^{n+\frac12}
    \left(1+a q^{n+1}z^m\right)
  \notag\\
  &=\exp\left[
    \sum_{r=1}^{\infty}\frac{(-1)^{r+1}}{r}
    a^r f(q^r,z^r)\right].
  \label{eq:majorana-plethystic}
\end{align}
The coefficient
\begin{equation}
  F_{2k}(q,z)=[a^{2k}]Z(a,q,z)
\end{equation}
counts operators containing $2k$ fermion fields, including conformal
descendants.  A generic primary has descendant character
\begin{equation}
  \mathcal D(q,z)
  =\frac{1}{(1-q)(1-qz)(1-q/z)}.
  \label{eq:majorana-descendants}
\end{equation}
Consequently, the signed primary numerator is
\begin{equation}
  \mathcal N_{2k}(q,z)
  =F_{2k}(q,z)(1-q)(1-qz)(1-q/z).
  \label{eq:majorana-primaries}
\end{equation}
At each power of $q$, decomposition into $\operatorname{Spin}(3)$ characters
gives the primary multiplicities.  Negative terms represent null descendants
of short multiplets.  In particular, a conserved current of spin $\ell$ and
dimension $\ell+1$ contributes a negative term of spin $\ell-1$ at dimension
$\ell+2$ for its vanishing divergence
\cite{Dolan:2005wy,Henning:2017fpj}.

\paragraph{Bilinear and higher fermion sectors}

The exterior square of the one fermion character is
\begin{equation}
  F_2(q,z)
  =\frac12\left[f(q,z)^2-f(q^2,z^2)\right].
  \label{eq:majorana-bilinear-closed}
\end{equation}
Substitution into Eq.~\eqref{eq:majorana-primaries} gives
\begin{equation}
  \mathcal N_2(q,z)
  =q^2\chi_0(z)
  +\sum_{\substack{\ell\geq2\\ \ell\ \mathrm{even}}}
   \left[q^{\ell+1}\chi_\ell(z)
   -q^{\ell+2}\chi_{\ell-1}(z)\right].
  \label{eq:majorana-bilinears}
\end{equation}
The scalar at dimension two is $\epsilon$.  The remaining positive terms are
the conserved currents $J_\ell$ of even spin, and the negative terms are their
null divergences.  There are no other bilinear primaries.

For four fermions, expansion of $F_4=[a^4]Z$ gives the primary multiplicities
shown in Table~\ref{tab:majorana-local-primary-count}.  The first six fermion
operator fills both level zero states and all four level one states.  It is a
scalar of dimension $2\cdot1+4\cdot2=10$.  No sector with eight or more
fermions occurs by dimension ten.

\begin{table}[H]
  \centering
  \begin{tabular}{c c l}
    \toprule
    sector & $\Delta$ & spin $\ell$ (multiplicity) \\
    \midrule
    $\psi^0$ & 0  & $0$ (identity) \\
    \addlinespace
    $\psi^2$ & 2  & $0$ ($\epsilon$) \\
    $\psi^2$ & 3  & $2$ ($T=J_2$) \\
    $\psi^2$ & 5  & $4$ ($J_4$) \\
    $\psi^2$ & 7  & $6$ ($J_6$) \\
    $\psi^2$ & 9  & $8$ ($J_8$) \\
    \addlinespace
    $\psi^4$ & 6  & $0,\ 2$ \\
    $\psi^4$ & 7  & $2,\ 4$ \\
    $\psi^4$ & 8  & $0,\ 2,\ 3,\ 4\ (2),\ 5$ \\
    $\psi^4$ & 9  & $2,\ 3,\ 4,\ 5,\ 6\ (2)$ \\
    $\psi^4$ & 10 & $0,\ 2\ (2),\ 4\ (3),\ 5\ (2),\ 6\ (3),
                         \ 7\ (2),\ 8$ \\
    \addlinespace
    $\psi^6$ & 10 & $0$ \\
    \bottomrule
  \end{tabular}
  \caption{Bosonic local conformal primaries in the free Majorana theory
  through $\Delta=10$, organized by fermion number.  Parentheses denote
  multiplicity.}
  \label{tab:majorana-local-primary-count}
\end{table}

The table counting local primaries and the OPE table answer different
questions.  Bose symmetry
removes odd spin from the OPE of identical scalars.  The bilinear scalar
$\epsilon$ is absent because it is odd under time reversal, while the even
spin currents are present.  In the four-fermion sector, the centered bilocal
\begin{equation}
  \mathcal B(x)=:\!\epsilon(x/2)\epsilon(-x/2)\!:
  =\mathcal B(-x)
  \label{eq:majorana-bilocal}
\end{equation}
contains only even powers of $x$, so only even dimensions occur.  Finally,
the sector decomposition in Eq.~\eqref{eq:majorana-exact-sector-sum} excludes
the six fermion scalar from this OPE.

These restrictions reduce the local spectrum below dimension ten precisely
to the sectors displayed in Table~\ref{tab:majorana-exact-ope}.  The only
degeneracy among them is at $(\Delta,\ell)=(8,4)$, where the local primary
space has dimension two.  The correlator fixes only the squared norm
$35/192$ of the OPE vector in this space.  One may choose an orthonormal basis
in which one primary carries this weight and the other has zero coefficient,
but the correlator alone does not select a preferred operator basis.  Every
other sector in Table~\ref{tab:majorana-exact-ope} contains a single local
primary.

\section{SDPB implementation and run parameters}
\label{app:sdpb-parameters}

This appendix records the complete numerical input used for the derivative
basis bounds.  Its purpose is to make the SDPB data reproducible without
requiring the reader to infer the settings at intermediate values of the
derivative order.

\subsection{Conformal block tables and reduced prefactors}

We consider identical external scalars in three dimensions, with
$\Delta_{12}=\Delta_{34}=0$.  Conformal block tables are generated with
\texttt{scalar\_blocks}\footnote{The source code is available at
\url{https://gitlab.com/bootstrapcollaboration/scalar_blocks}.} at odd
derivative order
\begin{equation}
  \Lambda=2n_{\max}-1,
  \qquad
  N_{\mathrm{der}}=\frac{n_{\max}(n_{\max}+1)}{2}.
\end{equation}
The complete production setup is labelled by
$(d,\Delta_{12},\Delta_{34},n_{\max},\texttt{lmax},\texttt{order},
\texttt{precision},n_{\mathrm{poles}})$.
These parameters enter at two different stages.  The conformal block cache
depends on the first seven entries and is independent of
$n_{\mathrm{poles}}$.
Here \texttt{lmax} is the largest spin, \texttt{order} is the truncation order
of the $r$ expansion, and \texttt{precision} is the working precision in
bits.  Following the terminology of Ref.~\cite{Chang:2025optimal}, the
\texttt{scalar\_blocks} parameter \texttt{poles} is denoted
\texttt{keptPoleOrder}.  In every block generation run we set
$\texttt{keptPoleOrder}=\texttt{order}$.  Consequently, no pole shifting is
performed in the block generation stage.

The parameter $n_{\mathrm{poles}}$ enters only when the polynomial matrix
program is converted to the SDP.  The production scripts call this input
\texttt{npoles}.  In the notation of Ref.~\cite{Chang:2025optimal},
\texttt{pmp2sdp} receives a damped rational \texttt{prefactor}
$\widetilde\mu^{(\ell)}(x)$ and a \texttt{reducedPrefactor}
$\mu^{(\ell)}(x)$ of the form
\begin{equation}
  \mu^{(\ell)}(x)
  =\frac{c_\ell r_\ell^x}
  {\prod_i (x-p_{\ell,i})^{\alpha_{\ell,i}}},
  \qquad
  \sum_i\alpha_{\ell,i}=n_{\mathrm{poles}}.
\end{equation}
The retained factors are the $n_{\mathrm{poles}}$ rightmost poles of
$\widetilde\mu^{(\ell)}(x)$, counted with multiplicity.  Thus
$n_{\mathrm{poles}}$ is the number of poles in the reduced prefactor.  It
controls the degree of the interpolating polynomials constructed by
\texttt{pmp2sdp}; it is neither a \texttt{scalar\_blocks} input nor a block
cache key.  The reduced prefactor also defines the weight used to construct
the optimal density interpolation nodes.

Because $\Delta_{12}=\Delta_{34}=0$, the same block cache can be reused for
every value of $\Delta_\phi$, either choice of the gap channel, and different
values of $n_{\mathrm{poles}}$.  The derivative count $N_{\mathrm{der}}$ is
not an additional numerical input; it follows from $\Lambda$ through the
formula above.

Table~\ref{tab:sdpb-runs} gives the complete production schedule.  Only odd
values of $\Lambda$ are used within each displayed range.

\begin{table}[H]
  \centering
  \small
  \begin{tabular}{ccccc}
    \toprule
    $\Lambda$ range & \texttt{lmax} & \texttt{order}
      & \texttt{precision} & $n_{\mathrm{poles}}$ \\
    \midrule
     $7$ to $11$ &  30 &  60 &  512 & 30 \\
    $13$ to $19$ &  46 &  80 &  512 & 40 \\
    $21$ to $35$ &  60 & 100 &  768 & 50 \\
    $37$ to $43$ &  80 & 120 &  768 & 60 \\
    $45$ to $63$ & 100 & 140 & 1024 & 80 \\
    \bottomrule
  \end{tabular}
  \caption{Block generation and SDP reduction schedule for the derivative
    basis runs.  The column names follow the production script:
    \texttt{lmax} is the largest spin, \texttt{order} is the truncation order
    of the $r$ expansion, and \texttt{precision} is the working precision in
    bits.  We set \texttt{keptPoleOrder} equal to \texttt{order}, while
    $n_{\mathrm{poles}}$ is the number of rightmost poles retained in the
    reduced prefactor used by \texttt{pmp2sdp}.  Each range contains only odd
    values of $\Lambda$.}
  \label{tab:sdpb-runs}
\end{table}

\subsection{SDPB settings and gap search}

For each trial gap, the Haskell driver, implemented with the Hyperion
framework,\footnote{The source code for Hyperion is available at
\url{https://github.com/davidsd/hyperion}.} assembles the crossing equations
into \texttt{continuum\_mat} JSON block matrices and writes an accompanying
\texttt{.nsv} file list.  The reduced prefactor described above is inserted
into these matrices immediately before \texttt{pmp2sdp} converts them to the
binary SDP representation.  The resulting problem is solved with
\textsc{SDPB}~3.1.0\footnote{The source code for \textsc{SDPB} is available at
\url{https://github.com/davidsd/sdpb}.}~\cite{Simmons-Duffin:2015qma}.

The boundary in the assumed gap is located by bisection until the final
interval has width below a certain threshold.  Every trial point is solved in
jump finding mode.  We disable \texttt{findPrimalFeasible} and
\texttt{findDualFeasible}, while enabling
\texttt{detectPrimalFeasibleJump} and \texttt{detectDualFeasibleJump}.  A
primal feasible jump classifies the trial gap as allowed, while a dual
feasible jump classifies it as excluded.  Thus the intended termination is a
feasibility jump, rather than convergence of the duality gap.  The primal and
dual error thresholds are fixed at $10^{-200}$ to avoid premature
classification near the boundary.

Table~\ref{tab:sdpb-params} gives the numerical scaling and termination
parameters for every run.  The working precision is supplied explicitly and
agrees with the precision used for the corresponding conformal block table.
Only odd values of $\Lambda$ in each displayed range are used.  All options
not listed in the table retain their \textsc{SDPB}~3.1.0 default values.

\begin{table}[H]
  \centering
  \scriptsize
  \setlength{\tabcolsep}{3.8pt}
  \begin{tabular}{@{}lccccccc@{}}
    \toprule
    $\Lambda$ range
      & $7$ to $11$ & $13$ to $19$ & $21$ to $27$ & $29$ to $35$
      & $37$ to $39$ & $41$ to $43$ & $45$ to $63$ \\
    \midrule
    precision (bits)
      & $512$ & $512$ & $768$ & $768$ & $768$ & $768$ & $1024$ \\
    \texttt{findPrimalFeasible}
      & \multicolumn{7}{c}{\texttt{False}} \\
    \texttt{findDualFeasible}
      & \multicolumn{7}{c}{\texttt{False}} \\
    \texttt{detectPrimalFeasibleJump}
      & \multicolumn{7}{c}{\texttt{True}} \\
    \texttt{detectDualFeasibleJump}
      & \multicolumn{7}{c}{\texttt{True}} \\
    \texttt{primalErrorThreshold}
      & \multicolumn{7}{c}{$10^{-200}$} \\
    \texttt{dualErrorThreshold}
      & \multicolumn{7}{c}{$10^{-200}$} \\
    \texttt{dualityGapThreshold}
      & $10^{-30}$ & $10^{-30}$ & $10^{-30}$ & $10^{-30}$
      & $10^{-30}$ & $10^{-75}$ & $10^{-75}$ \\
    \texttt{initialMatrixScalePrimal}
      & $10^{20}$ & $10^{40}$ & $10^{50}$ & $10^{50}$
      & $10^{60}$ & $10^{60}$ & $10^{60}$ \\
    \texttt{initialMatrixScaleDual}
      & $10^{20}$ & $10^{40}$ & $10^{50}$ & $10^{50}$
      & $10^{60}$ & $10^{60}$ & $10^{60}$ \\
    \texttt{maxComplementarity}
      & $10^{100}$ & $10^{100}$ & $10^{130}$ & $10^{160}$
      & $10^{180}$ & $10^{200}$ & $10^{200}$ \\
    \bottomrule
  \end{tabular}
  \caption{SDPB solver parameters as a piecewise function of the derivative
    order.  The precision agrees with that used to generate the corresponding
    conformal block table.  Columns refer only to the odd derivative orders in
    the displayed ranges.}
  \label{tab:sdpb-params}
\end{table}

For spectrum extraction, we run the \texttt{spectrum} utility distributed with
\textsc{SDPB}~3.1.0, using the same precision as the corresponding solver
run.  The zero threshold is $10^{-4}$.

\section{Numerical implementation of analytic functionals}
\label{app:analytic-numerics}

%The numerical calculation separates the construction of the functional actions from the subsequent bootstrap optimization.  
Our numerical implementation proceeds in two steps: we first generate the table of functional actions and then use it in the linear programming. The open source Wolfram
Language package
\href{https://github.com/Canonical111/FunTable}{\texttt{FunTable}} evaluates
the actions at fixed external dimension and stores them as HDF5 tables.  The
Julia package
\href{https://github.com/Canonical111/FunBoot}{\texttt{FunBoot}} reads
these tables, interpolates in the exchanged dimension, and solves the
resulting linear program.  This organization allows the expensive arbitrary-precision evaluation of the functionals to be performed once and reused in
different bootstrap problems.  The optimization framework builds on the
solver used in \cite{Ghosh:2023onl}.

\subsection{Generation of the functional tables}

At functional order $\lambda$, the numerical basis contains
\begin{equation}
  \boxed{N_{\mathrm{fun}}=2(\lambda+1)^2}
  \label{eq:analytic-basis-size}
\end{equation}
functional components.   In
particular, $\lambda=12$ gives $N_{\mathrm{fun}}=338$.

For a three-dimensional correlator with external dimension $\Delta_\phi$,
\texttt{FunTable} first evaluates the one-dimensional bosonic and fermionic
functionals at external dimension $\Delta_\phi/2$.  Their actions are
generated on the grid
\begin{equation}
  \Delta^{1d}_k=\frac{k}{202}.
\end{equation}
A recursion in the functional index is initialized with regularized
hypergeometric functions.  Direct arbitrary-precision evaluation is used at
low dimension.  At large dimension it is matched smoothly to the asymptotic
expansion described in \smref{app:asympt}.  The working precision is
increased dynamically until a sample at the matching point retains at least
$27$ significant decimal digits.

The one-dimensional actions are combined into two-dimensional product
functionals and uplifted to three dimensions through dimensional reduction.
For an operator of even spin $\ell$, the action takes the form
\begin{equation}
(\omega^-_m\otimes\omega^+_n)
\!\left[
F^{d=3}_{\Delta,\ell}
\right]
=
\sum_{q=0}^{\infty}
\sum_{s=\ell,\ell-2,\ldots,\ell\,{\rm mod}\,2}
A_{q,s}(\Delta,\ell)\,
(\omega^-_m\otimes\omega^+_n)
\bigl[F^{d=2}_{\Delta+2q,s}\bigr].
\label{eq:3dfunc2}
\end{equation}
The sum over $q$ is evaluated directly through $N_{\mathrm{red}}$.  Its
remaining tail is expressed in terms of Hurwitz zeta functions.  In the
production tables, the estimated tail near the unitarity bound is of order
$10^{-19}$.

Before the numerical values are stored, the one-dimensional actions are
rescaled by $2^{-2\Delta}$.  This removes their exponential growth and avoids
overflow at large $\Delta$.  The final tables contain all even spins from
$0$ to $100$ on the uniform three-dimensional grid
\begin{equation}
  \Delta_k=\frac{k}{101},
  \qquad
  0\leq k\leq15150.
\end{equation}
Thus every spin table extends to $\Delta=150$.  The HDF5 metadata record
$\Delta_\phi$, $\lambda$, the basis size, the spin list, the grid,
$N_{\mathrm{red}}$, and the generation precision.  At the highest functional
order used in this work, the generation parameters are
\begin{equation}
  \lambda=12,
  \qquad
  N_{\mathrm{fun}}=338,
  \qquad
  \Delta_{\max}^{1d}=650,
  \qquad
  N_{\mathrm{red}}=500.
  \label{eq:analytic-table-parameters}
\end{equation}
Here $\Delta_{\max}^{1d}$ is the endpoint of the one-dimensional generation
grid.  The value of $N_{\mathrm{fun}}$ follows from
Eq.~\eqref{eq:analytic-basis-size}.

\subsection{Interpolation and optimization}

The tabulated actions are interpolated in $\Delta$ with local barycentric
Lagrange polynomials.  The degree is $12$ for $\lambda\leq5$ and $28$ for
$\lambda>5$.
  We implement these routines in
\href{https://github.com/Canonical111/LocalBarycentric.jl}
{\texttt{LocalBarycentric.jl}}, a registered Julia package developed by the
authors.  The results were also checked against an independent Wolfram
Language interpolation of the same tables.

\texttt{FunBoot} formulates gap maximization as a linear program and solves
it with the simplex method inside an outer approximation procedure
\cite{Ghosh:2023onl}.  The initial candidate set samples the target spin at
full grid density and every third grid point in the other spin sectors.
Candidates within $10^{-10}$ of
$\Delta-\ell=2\Delta_\phi+2n$ are omitted because the functional actions have
asymptotic double zeros there.  Following Ref.~\cite{Paulos:2021jxx}, we
assign an exponentially decaying cost to the target sector.  For a gap in the
sector of spin
$\ell_{\mathrm{gap}}$, the initial cost is
\begin{equation}
  \operatorname{cost}(\Delta,\ell)
  =
  \begin{cases}
    10^{220-c(\Delta-\ell)}, & \ell=\ell_{\mathrm{gap}},\\
    0, & \ell\neq\ell_{\mathrm{gap}},
  \end{cases}
  \qquad c=10.
  \label{eq:analytic-cost}
\end{equation}
Minimizing this cost pushes the first operator in the target sector toward
the largest dimension compatible with crossing and positivity.
More broadly, using a continuous objective to guide a feasibility search is
similar in spirit to the navigator function \cite{Reehorst:2021ykw}, although here the optimization proceeds through
a sequence of linear programs within an outer approximation procedure.

After the initial solve, new candidate operators are inserted near violations
of positivity and near the active operators.  At refinement step $k$, the
cost parameter is $c_k=10\,3^{k-1}$.  The production runs use twelve
refinement steps within $\ell<60$ and $\Delta<100$.  Spins outside this region
remain as asymptotic constraints.  The remaining settings are summarized in
Table~\ref{tab:analytic-optimization-settings}.

\begin{table}[H]
  \centering
  \begin{tabular}{p{0.46\linewidth}p{0.43\linewidth}}
    \toprule
    parameter & value \\
    \midrule
    standard spin set
      & $\ell=0,2,\ldots,60$, together with $\ell=80$ \\
    spin set at large $\lambda$
      & $\ell=0,2,\ldots,80$, together with $\ell=100$ \\
    simplex iteration limit
      & $20000$ initially and $10000$ at each refinement step \\
    duplicate candidate tolerance
      & relative dimension difference $10^{-9}$ \\
    iterative refinement
      & one step at fixed precision and five steps at mixed precision \\
    refinement domain
      & $\ell<60$ and $\Delta<100$ \\
    \bottomrule
  \end{tabular}
  \caption{Numerical parameters used in the analytic-functional optimization.
  Parameters relevant only to diagnostics and output formatting are omitted.}
  \label{tab:analytic-optimization-settings}
\end{table}

The simplex calculation begins in standard machine precision.  Its numerical
stability is monitored through the basis indicator
\begin{equation}
  \iota=\left\|\mathbf{1}-\widehat{B^{-1}}B\right\|,
  \label{eq:basis-indicator}
\end{equation}
where $\widehat{B^{-1}}$ is the computed inverse of the current simplex basis.
When $\iota>10^{-2}$, the basis algebra is repeated with
\texttt{Double64} arithmetic from \texttt{DoubleFloats.jl}.  Stable
\texttt{Double64} runs typically give $\iota\sim10^{-18}$.  The final
objective and residuals are checked independently with $256$ bit
\texttt{BigFloat} arithmetic.

\subsection{Normalization and spectrum extraction}

The functional tables used in the numerical implementation are normalized by factoring out a suitable dimension- and spin-dependent prefactor, chosen to control the large-dimension behavior of the product functionals. So the squared OPE coefficients returned by the linear program are converted to the physical
three dimensional normalization according to
\begin{equation}
  f^{2}_{\mathrm{phys}}
  =
  f^{2}_{\mathrm{LP}}\,4^{-\Delta}
  \left(1+\frac{\Delta+\ell}{2}\right)^{2\Delta_\phi+\frac32}
  \left(1+\frac{\Delta-\ell}{2}\right)^{2\Delta_\phi+\frac72}
  \frac{(1/2)_\ell}{\ell!}\frac{1}{\sqrt{1+\Delta}}.
  \label{eq:analytic-ope-normalization}
\end{equation}
The factor $(1/2)_\ell/\ell!$ implements the unit normalized conformal block
convention used in this work.

Their squared OPE coefficients are added, and the
reported dimension is their OPE weighted mean.  Operators with $\ell\geq2$
that lie within $10^{-4}$ of the conserved locus $\Delta=\ell+1$ are clustered
only with other operators satisfying the same condition.  This prescription
keeps conserved currents distinct from nearby unprotected operators.

Both public repositories contain worked examples and documentation.  The
\texttt{FunBoot} repository also pins the Julia environment, includes
regression tests, and supports deterministic checkpointing of long runs.
These features provide the complete numerical path from the analytic
functional definitions to the bounds and extremal spectra reported here.

\section{Numerical results and convergence diagnostics}
\label{app:funconv}

In this appendix, we collect the numerical output underlying the bounds and
extremal spectra discussed in the main text.  The first two subsections present
the complete scans in the external dimension.  The next two examine
convergence at the Ising point and at $\Delta_\phi=2$.  The final subsection
records the complete derivative basis data used in these comparisons.

\subsection{Complete gap scans}
\begin{figure}[tbp]
  \centering
  \includegraphics[width=0.84\linewidth]{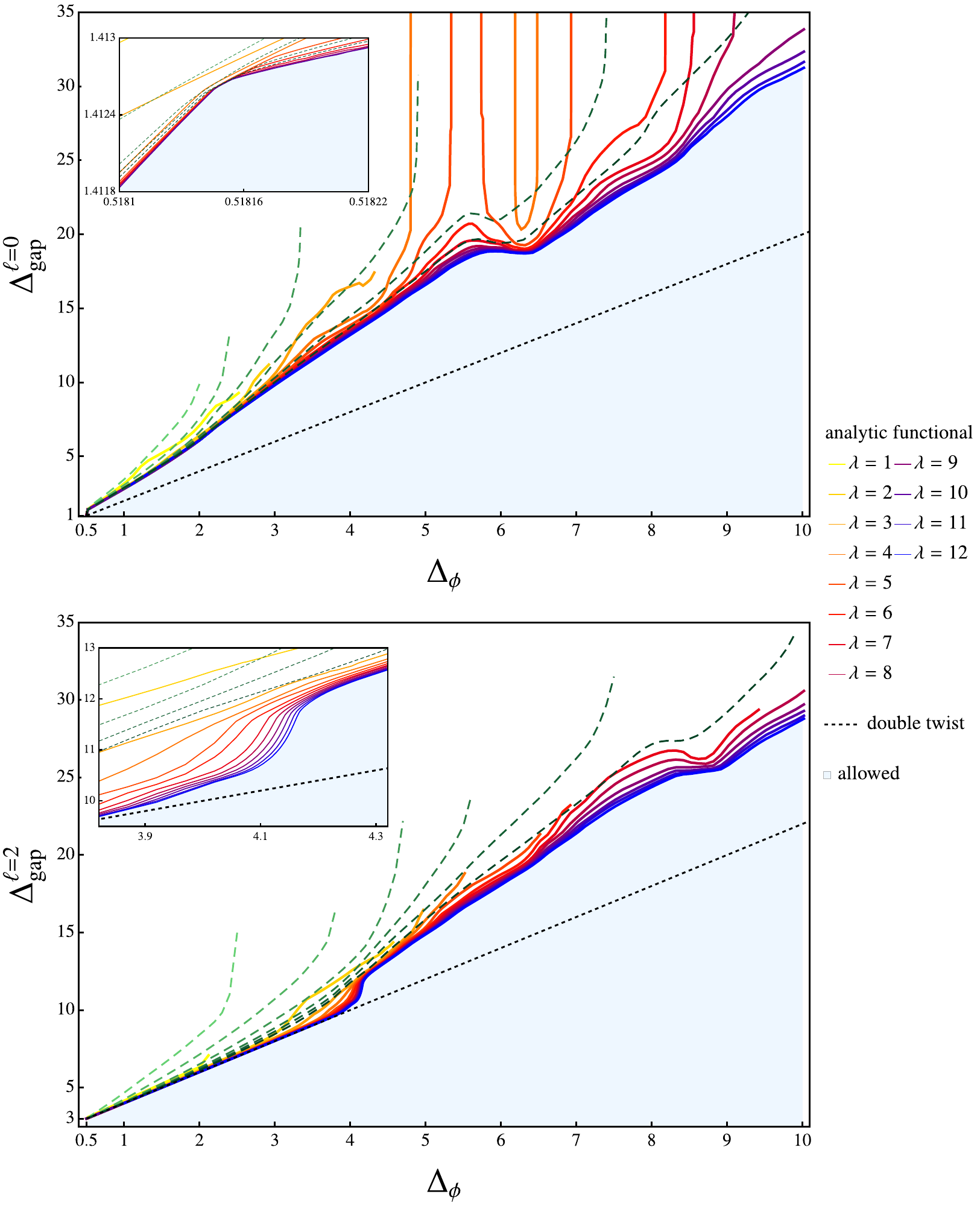}
  \caption{%
    Complete cutoff comparison for the single-correlator gap maximization
    problems in $d=3$.  The upper panel shows the scalar gap, and the lower
    panel shows the gap for operators of spin two without an isolated stress
    tensor.
    Solid curves give the analytic functional results for $\lambda=1$ to
    $12$.  Dashed green curves give the derivative basis results for
    $\Lambda\in\{7,11,19,29,43,63\}$.  Darker curves correspond to larger
    cutoffs.  The black dotted line is the leading GFF double-twist value.
    The insets show the Ising region in the upper panel and the kink region
    in the lower panel.%
  }
  \label{fig:gap-benchmark}
\end{figure}
Figure~\ref{fig:gap-benchmark} shows all available cutoffs for both functional bases and complements the more selective comparison in \mainref{fig:gap-bounds}. This makes the convergence pattern and the finite-cutoff artifacts visible across the full range of $\Delta_\phi$. At fixed $\Delta_\phi$, increasing the cutoff lowers the exclusion curve toward the limiting bound. The analytic-functional curves are already smooth at the largest cutoffs, while the low-order derivative curves become loose at large $\Delta_\phi$, producing steep segments that disappear as $\Lambda$ increases. The inset of the lower panel also shows the stabilization of the sharp feature near $\Delta_\phi\simeq4.16$. The upper and lower panels develop broad plateaus near $\Delta_\phi\simeq6.5$ and $\Delta_\phi\simeq8.5$, respectively, where successive cutoff curves approach one another more rapidly than in the surrounding regions. This increased stability suggests that these plateaus may be associated with distinguished crossing solutions, although convergence alone is not enough to establish a CFT interpretation. They therefore deserve a more detailed study of the corresponding extremal spectra.

\subsection{Extremal spectra across the external dimension}

\begin{figure}[tbp]
  \centering
  \includegraphics[width=0.84\linewidth]{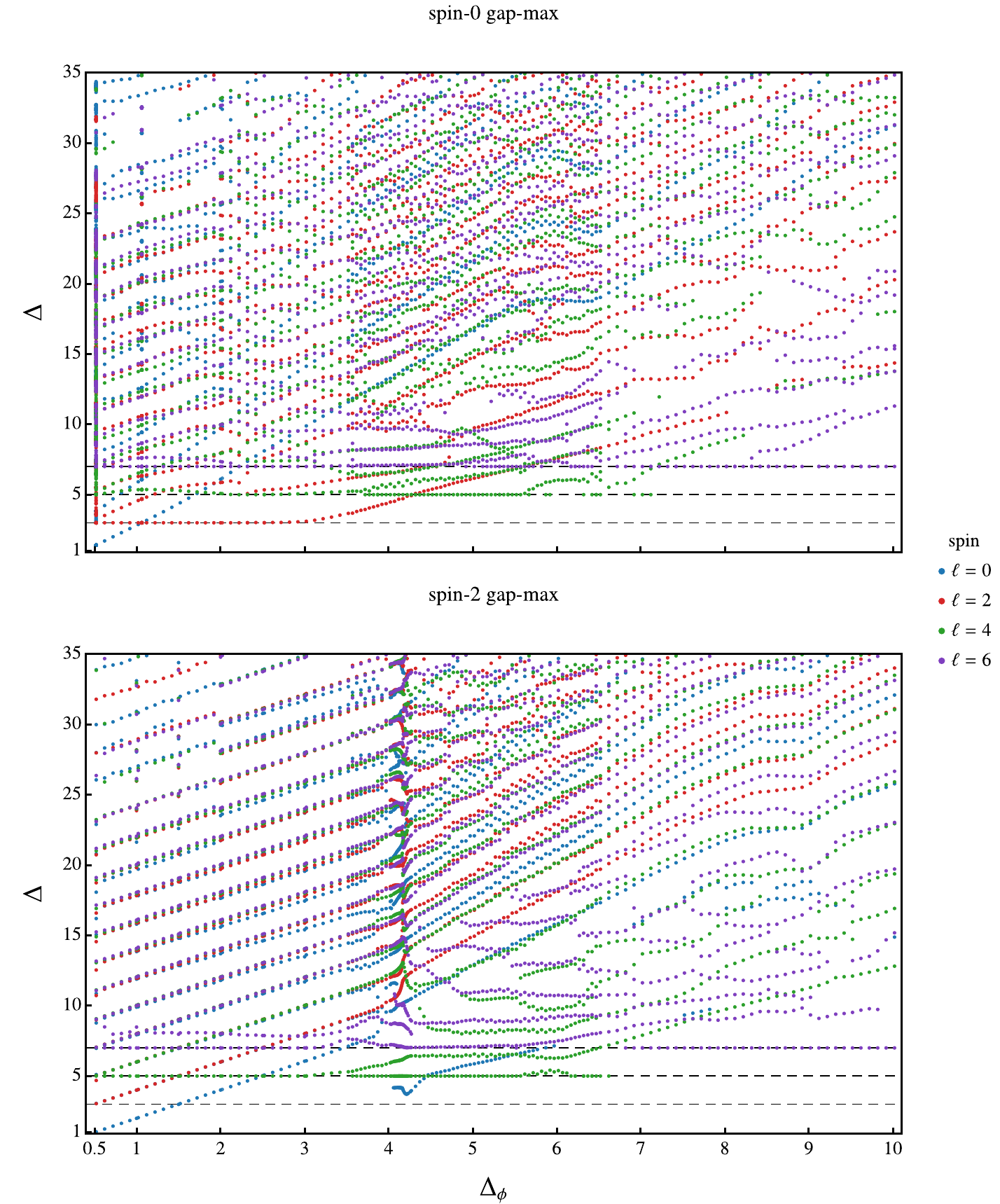}
  \caption{%
    Extremal spectra along the full external dimension scan.  The spectra
    use $\lambda=12$, except for $\Delta_\phi<2$, where they use
    $\lambda=10$.  The points at $\Delta_\phi=1$ and $1.5$ are exceptions
    and use $\lambda=12$.  The upper panel corresponds to scalar gap
    maximization, and the lower panel corresponds to gap maximization for
    operators of spin two.  Marker color denotes spin.  The dense column in
    the lower panel is the fine scan around the spectral reorganization near
    $\Delta_\phi\simeq4.16$.%
  }
  \label{fig:spectrum-dphi}
\end{figure}

%For each point on the analytic bound, 
At each point on the bound obtained with the analytic-functional basis, we read off the extremal spectrum
from the support of the optimal primal solution.  The result is shown in
Fig.~\ref{fig:spectrum-dphi}. 
In the lower panel, the spectrum begins to lose its GFF double-twist
organization for $\Delta_\phi\gtrsim3$, even while the leading gap remains close to
$2\Delta_\phi+2$.  The much sharper reorganization near
$\Delta_\phi\simeq4.16$ is analyzed in \smsecref{sec:gff}.  A stress tensor
at $(\Delta,\ell)=(3,2)$ is absent from the lower panel, since the assumed gap
excludes it.

\subsection{Convergence at the Ising point}

Table~\ref{tab:ising_l0} gives convergence of the
leading scalar dimension and OPE coefficient at the Ising point using the
product analytic functional basis.
Table~\ref{tab:ising-eps-bound} gives the corresponding derivative basis
bound.  The first column is obtained directly from
gap bisection.  The remaining columns are zeros of the optimal functional,
extracted with the SDPB spectrum threshold $10^{-4}$.  A missing value means
that no zero was resolved inside the tracking window at that cutoff.

\begin{table}[tbp]
  \centering
  \begin{tabular}{r l l l}
    \toprule
    $\lambda$ & $\Delta_{\epsilon}(\lambda)$ & $\Delta_{\epsilon}-\Delta_{\epsilon}^{\mathrm{ref}}$ & $f_{\sigma\sigma\epsilon}^{2}$ \\
    \midrule
     1 & 1.4436410201 & $3.10\times10^{-2}$ & 1.0758441560 \\
     2 & 1.4133371669 & $7.12\times10^{-4}$ & 1.1056881030 \\
     3 & 1.4127680141 & $1.43\times10^{-4}$ & 1.1062579050 \\
     4 & 1.4126700862 & $4.51\times10^{-5}$ & 1.1063516460 \\
     5 & 1.4126389395 & $1.39\times10^{-5}$ & 1.1063826010 \\
     6 & 1.4126349333 & $9.93\times10^{-6}$ & 1.1063865840 \\
     7 & 1.4126329977 & $8.00\times10^{-6}$ & 1.1063885280 \\
     8 & 1.4126321207 & $7.12\times10^{-6}$ & 1.1063894120 \\
     9 & 1.4126315183 & $6.52\times10^{-6}$ & 1.1063900150 \\
    10 & 1.4126312814 & $6.28\times10^{-6}$ & 1.1063902530 \\
    \midrule
    ref. & 1.412625 & --- & 1.106396 \\
    \bottomrule
  \end{tabular}
  \caption{Leading $\ell=0$ scalar $\epsilon$ at the $3$d Ising point
  $\Delta_\phi=0.5181488023$ from $\ell=0$ gap maximization at analytic cutoff
  $\lambda$: dimension $\Delta_\epsilon(\lambda)$, its deviation from the reference,
  and the extremal OPE coefficient $f^2_{\sigma\sigma\epsilon}$.  Reference values
  $\Delta_\epsilon=1.412625$, $f^2_{\sigma\sigma\epsilon}\approx1.106396$
  ($f_{\sigma\sigma\epsilon}=1.0518537$), Kos--Poland--Simmons-Duffin--Vichi
  \cite{Kos:2016ysd}.}
  \label{tab:ising_l0}
\end{table}

\begin{table}[b!]
  \centering
  \footnotesize
  \renewcommand{\arraystretch}{1.05}
  \setlength{\tabcolsep}{4pt}
  \begin{tabular}{r|lll|l@{\hspace{2em}}r|lll|l}
    \hline\hline
     & \multicolumn{3}{c|}{spin $0$} & spin $2$ &
     & \multicolumn{3}{c|}{spin $0$} & spin $2$ \\
    $\Lambda$ & $\Delta_{\epsilon}^{\,\ast}$ & $\Delta_{\epsilon'}$ & $\Delta_{\epsilon''}$ & $\Delta_{T'}$ &
    $\Lambda$ & $\Delta_{\epsilon}^{\,\ast}$ & $\Delta_{\epsilon'}$ & $\Delta_{\epsilon''}$ & $\Delta_{T'}$ \\
    \hline
    7  & $1.428825867410(7)$  & --        & --        & --        & 37 & $1.4126485938(4)$ & $3.83029$ & $7.01570$ & $5.51163$ \\
    9  & $1.4205058965(3)$    & $4.11227$ & --        & $5.09414$ & 39 & $1.4126436887(4)$ & $3.83015$ & $6.99806$ & $5.51112$ \\
    11 & $1.417843014811(8)$  & $4.02500$ & --        & $5.13736$ & 41 & $1.4126414478(4)$ & $3.83009$ & $6.98945$ & $5.51094$ \\
    13 & $1.4138394376(3)$    & $3.86640$ & --        & $5.48105$ & 43 & $1.4126391440(5)$ & $3.83002$ & $6.98051$ & $5.51071$ \\
    15 & $1.4134181951(3)$    & $3.85552$ & --        & $5.46714$ & 45 & $1.4126378494(4)$ & $3.82999$ & $6.97535$ & $5.51060$ \\
    17 & $1.4128697715(3)$    & $3.83698$ & --        & $5.51470$ & 47 & $1.4126367581(4)$ & $3.82996$ & $6.97075$ & $5.51051$ \\
    19 & $1.41280960314(1)$   & $3.83549$ & $7.64103$ & $5.51064$ & 49 & $1.4126358667(4)$ & $3.82993$ & $6.96721$ & $5.51040$ \\
    21 & $1.4127201461(3)$    & $3.83237$ & $7.29802$ & $5.51563$ & 51 & $1.4126351841(4)$ & $3.82991$ & $6.96430$ & $5.51034$ \\
    23 & $1.4127059238(3)$    & $3.83196$ & $7.23356$ & $5.51485$ & 53 & $1.4126346846(4)$ & $3.82990$ & $6.96230$ & $5.51028$ \\
    25 & $1.4126855672(3)$    & $3.83132$ & $7.14497$ & $5.51445$ & 55 & $1.4126342656(4)$ & $3.82988$ & $6.96048$ & $5.51024$ \\
    27 & $1.4126804805(3)$    & $3.83117$ & $7.12331$ & $5.51414$ & 57 & $1.4126338372(4)$ & $3.82987$ & $6.95860$ & $5.51021$ \\
    29 & $1.41267217019(1)$   & $3.83093$ & $7.09188$ & $5.51366$ & 59 & $1.4126335772(4)$ & $3.82986$ & $6.95753$ & $5.51017$ \\
    31 & $1.4126680925(4)$    & $3.83083$ & $7.08151$ & $5.51326$ & 61 & $1.4126332908(4)$ & $3.82986$ & $6.95608$ & $5.51018$ \\
    33 & $1.4126610709(4)$    & $3.83063$ & $7.05748$ & $5.51275$ & 63 & $1.4126331071(1)$ & $3.82985$ & $6.95527$ & $5.51016$ \\
    35 & $1.4126522443(4)$    & $3.83039$ & $7.02864$ & $5.51193$ &    &                   &           &           &           \\
    \hline\hline
  \end{tabular}
  \caption{Leading scalar bound and subleading extremal functional zeros at
    the Ising external dimension, as functions of
    $\Lambda=2n_{\max}-1$.  The spin-$0$ columns contain the scalars even
    under $\mathbb Z_2$.}
  \label{tab:ising-eps-bound}
\end{table}

The leading scalar bracket has width below $10^{-9}$.  The subleading
dimensions are quoted only to show their convergence and should be assigned
an uncertainty of order $10^{-3}$.  For comparison, mixed-correlator studies
give $\Delta_\epsilon\simeq1.412625$,
$\Delta_{\epsilon'}\simeq3.82968$,
$\Delta_{\epsilon''}\simeq6.8956$, and
$\Delta_{T'}\simeq5.50915$~\cite{Simmons-Duffin:2016wlq}.  At the largest
cutoff, the leading scalar, $\epsilon'$, and $T'$ are already close to
these values, whereas $\epsilon''$ retains the displacement discussed in
\smsecref{sec:ising-kink}.

\subsection{Complete derivative basis data}

Tables~\ref{tab:sdpb-scalar-dphi2} and \ref{tab:sdpb-bounds} record the
complete derivative basis sequences used in the convergence comparisons.
The first table contains the scalar gap at $\Delta_\phi=2$, while the second
contains the gaps for operators of spin two at five external dimensions.
Each entry is the midpoint of the final allowed and excluded bisection
bracket, and the digit in parentheses is half the bracket width.  All final
brackets have width below $10^{-6}$.  At $\Delta_\phi=3$ and $\Lambda=7$, no
exclusion was found below the search ceiling $\Delta=60$.

\begin{table}[tbp]
  \centering
  \small
  \renewcommand{\arraystretch}{1.08}
  \begin{tabular}{rc@{\qquad}rc@{\qquad}rc}
    \toprule
    $\Lambda$ & $\Delta_{\ell=0}^{*}$
      & $\Lambda$ & $\Delta_{\ell=0}^{*}$
      & $\Lambda$ & $\Delta_{\ell=0}^{*}$ \\
    \midrule
     7 & $9.8945334(5)$ & 27 & $6.4374011(3)$ & 47 & $6.1848225(3)$ \\
     9 & $8.5915941(3)$ & 29 & $6.3652371(5)$ & 49 & $6.1777249(3)$ \\
    11 & $7.5831819(5)$ & 31 & $6.3313579(4)$ & 51 & $6.1736876(3)$ \\
    13 & $7.0701138(5)$ & 33 & $6.2877191(4)$ & 53 & $6.1686707(3)$ \\
    15 & $6.8568835(5)$ & 35 & $6.2676539(4)$ & 55 & $6.1661129(3)$ \\
    17 & $6.6888228(5)$ & 37 & $6.2404297(4)$ & 57 & $6.1617571(3)$ \\
    19 & $6.6453935(5)$ & 39 & $6.2272752(4)$ & 59 & $6.1599767(3)$ \\
    21 & $6.5565067(3)$ & 41 & $6.2107551(4)$ & 61 & $6.1568331(3)$ \\
    23 & $6.5453020(3)$ & 43 & $6.2016353(5)$ & 63 & $6.1552796(5)$ \\
    25 & $6.4679545(3)$ & 45 & $6.1903146(3)$ &    &                \\
    \bottomrule
  \end{tabular}
  \caption{Complete derivative basis sequence for the scalar gap at
    $\Delta_\phi=2$, as a function of $\Lambda=2n_{\max}-1$.}
  \label{tab:sdpb-scalar-dphi2}
\end{table}

\begin{table}[tbp]
  \centering
  \scriptsize
  \renewcommand{\arraystretch}{1.06}
  \begin{tabular}{rccccc}
    \toprule
    & \multicolumn{5}{c}{Gap for operators of spin two,
      $\Delta_{\ell=2}^{*}$} \\
    \cmidrule(lr){2-6}
    $\Lambda$ & $\Delta_\phi=1$ & $1.5$ & $2$ & $2.5$ & $3$ \\
    \midrule
     7 & $4.6696114(4)$ & $6.4612500(4)$ & $8.412399301(9)$ & $14.9948338(4)$ & $\cdots$ \\
     9 & $4.3015854(4)$ & $5.7672372(4)$ & $7.2911216(3)$ & $8.9428288(4)$ & $11.2295566(3)$ \\
    11 & $4.2517302(3)$ & $5.6418880(3)$ & $7.13725062(4)$ & $8.8013071(5)$ & $10.7130074(4)$ \\
    13 & $4.1451881(3)$ & $5.4084554(4)$ & $6.7900442(3)$ & $8.3755122(4)$ & $10.0132264(3)$ \\
    15 & $4.1350613(3)$ & $5.3802974(4)$ & $6.7355517(3)$ & $8.2025238(4)$ & $9.8107509(3)$ \\
    17 & $4.0894393(3)$ & $5.2850587(4)$ & $6.5672183(3)$ & $7.9346016(4)$ & $9.5094366(3)$ \\
    19 & $4.0860809(5)$ & $5.2627813(3)$ & $6.51527582(7)$ & $7.8766869(4)$ & $9.3787819(3)$ \\
    21 & $4.0635920(4)$ & $5.2026599(3)$ & $6.4307585(4)$ & $7.7529914(3)$ & $9.1735770(3)$ \\
    23 & $4.0574830(4)$ & $5.1907607(3)$ & $6.4035214(4)$ & $7.6977379(3)$ & $9.0949817(3)$ \\
    25 & $4.0463853(4)$ & $5.1622679(3)$ & $6.3459820(4)$ & $7.6134932(3)$ & $8.9781537(3)$ \\
    27 & $4.0434090(4)$ & $5.1522299(3)$ & $6.3240856(4)$ & $7.5744977(3)$ & $8.9197201(3)$ \\
    29 & $4.0364115(3)$ & $5.1311073(3)$ & $6.28789775(6)$ & $7.5225080(4)$ & $8.8483363(3)$ \\
    31 & $4.0339379(2)$ & $5.1233632(4)$ & $6.2722038(4)$ & $7.4957427(3)$ & $8.8109259(4)$ \\
    33 & $4.0294058(2)$ & $5.1106695(4)$ & $6.2490174(4)$ & $7.4605784(3)$ & $8.7644277(4)$ \\
    35 & $4.0277435(2)$ & $5.1045343(4)$ & $6.2375669(4)$ & $7.4411885(3)$ & $8.7364924(4)$ \\
    37 & $4.0248481(2)$ & $5.0959243(4)$ & $6.2206395(4)$ & $7.4155701(3)$ & $8.6994924(4)$ \\
    39 & $4.0236734(2)$ & $5.0917122(4)$ & $6.2119229(4)$ & $7.4002005(3)$ & $8.6762344(4)$ \\
    41 & $4.0216847(2)$ & $5.0854486(4)$ & $6.1986237(4)$ & $7.3778288(3)$ & $8.6446389(4)$ \\
    43 & $4.0208007(2)$ & $5.0823333(4)$ & $6.19148247(7)$ & $7.3638668(3)$ & $8.6271368(4)$ \\
    45 & $4.0193810(4)$ & $5.0769895(3)$ & $6.1802236(5)$ & $7.3451110(3)$ & $8.6017236(3)$ \\
    47 & $4.0186522(4)$ & $5.0742264(3)$ & $6.1742357(5)$ & $7.3347338(3)$ & $8.5858510(3)$ \\
    49 & $4.0174792(4)$ & $5.0698902(3)$ & $6.1649442(5)$ & $7.3196068(3)$ & $8.5638194(3)$ \\
    51 & $4.0168966(4)$ & $5.0677303(3)$ & $6.1604454(5)$ & $7.3106763(3)$ & $8.5497655(3)$ \\
    53 & $4.0159105(4)$ & $5.0642809(3)$ & $6.1526200(5)$ & $7.2960885(3)$ & $8.5263362(3)$ \\
    55 & $4.0154669(4)$ & $5.0625601(3)$ & $6.1487521(5)$ & $7.2884379(3)$ & $8.5150776(3)$ \\
    57 & $4.0146787(4)$ & $5.0594784(3)$ & $6.1415043(5)$ & $7.2748946(3)$ & $8.4928667(3)$ \\
    59 & $4.0143010(4)$ & $5.0579249(3)$ & $6.1381509(5)$ & $7.2690873(3)$ & $8.4822413(3)$ \\
    61 & $4.0135830(4)$ & $5.0551001(3)$ & $6.1316457(5)$ & $7.2576763(3)$ & $8.4635491(3)$ \\
    63 & $4.0132620(3)$ & $5.0539132(3)$ & $6.1289891(4)$ & $7.2521913(3)$ & $8.4546842(3)$ \\
    \bottomrule
  \end{tabular}
  \caption{Complete derivative basis sequences for the gaps of operators of
    spin two at the indicated external dimensions, as functions of
    $\Lambda=2n_{\max}-1$.}
  \label{tab:sdpb-bounds}
\end{table}

\FloatBarrier

\bibliographystyle{jhep}
\bibliography{ref}

\end{document}